\documentclass[prd, aps, nofootinbib, preprintnumbers, showpacs,
superscriptaddress,twocolumn, 10pt, floatfix]{revtex4-1}
\pdfoutput=1
\usepackage{graphics,graphicx,rotating}
\usepackage[usenames]{color}
\usepackage[percent]{overpic}
\usepackage{mathrsfs,amsmath,amssymb}
\usepackage{times}
\usepackage{mathptmx}
\usepackage{bm,multirow}

\newcommand{\beq}{\begin{equation}}
\newcommand{\eeq}{\end{equation}}

\newcommand{\AEI}{\affiliation{Max-Planck-Institut f\"ur
    Gravitationsphysik, Am M\"uhlenberg 1, 14475 Potsdam, Germany}}

\newcommand{\palma}{\affiliation{Departament de F\'isica, 
  Universitat de les Illes Balears, Cra.\ Valldemossa Km.\ 7.5, Palma 
de Mallorca, E-07122 Spain}}

\newcommand{\cardiff}{\affiliation{School of Physics and Astronomy, Cardiff
University, Cardiff, CF24 3AA, United Kingdom}}

\newcommand{\invisible}[1]{}

\begin{document}

\title{Reliability of complete gravitational waveform models for
compact binary coalescences}

\author{Frank Ohme}         \AEI
\author{Mark Hannam}        \cardiff
\author{Sascha Husa}        \palma

\date{September 12, 2011}

\begin{abstract}
Accurate knowledge of the gravitational-wave (GW) signal from
inspiraling compact binaries is essential to detect these signatures
in the data from GW interferometers. With recent
advances in post-Newtonian (PN) theory and numerical relativity (NR)
it has become possible to construct inspiral-merger-ringdown waveforms
by combining both descriptions into one complete \emph{hybrid} signal.
While addressing the reliability of such waveforms in different
points of the physical parameter space, previous studies have
identified the PN contribution as the dominant source of error, which
can be reduced by incorporating longer NR simulations. 
In this paper we overcome the 
two outstanding issues that make it difficult to determine the minimum
simulation length necessary to produce suitably accurate hybrids for
GW astronomy applications: (1) the relevant criteria for a GW search
is the mismatch between the true waveform and a set of model
waveforms, optimized over all waveforms in the model, but for discrete
hybrids this optimization was not yet possible. (2) these
calculations typically require that numerical waveforms already exist,
while we develop an algorithm to estimate hybrid
mismatch errors \emph{without} numerical data, which enables us
to estimate the necessary NR waveform length \emph{before} performing the
simulation. 
Our procedure relies on combining supposedly equivalent PN models at
highest available order with common data in the NR regime, and their
difference serves as a measure of the uncertainty assumed in each
waveform.
Contrary to some earlier studies, we
estimate that $\sim$10 NR orbits before merger should allow for the
construction of waveform families that are accurate enough for
detection in a broad range of parameters, only excluding highly
spinning, unequal-mass systems. Nonspinning systems, even with high
mass-ratio ($q \gtrsim 20$) are well modeled for astrophysically
reasonable component masses. In addition, the parameter bias is only
of the order of 1\% for total mass and symmetric mass-ratio and less
than 0.1 for the dimensionless spin magnitude.  We take the view that
similar NR waveform lengths will remain the state of the art in the
advanced detector era, and begin to assess the limits of the science
that can be done with them. 
\end{abstract}


\preprint{LIGO-P1100078-v3/AEI-2011-039}

\maketitle

\section{Introduction}

A network of gravitational-wave (GW) detectors is preparing 
to achieve a remarkable scientific goal: the first direct detection 
of GWs. This will not only test the predictions from Einstein's general 
theory of relativity, it will also open a new
window to the universe, revealing details of the population,
composition and formation history of various astrophysical 
objects~\cite{Sathyaprakash:2009xs}.
One particularly interesting and
promising source of detectable GWs is the inspiral, merger and
ringdown of compact objects, such as black holes or neutron stars. 

An important contribution to the effort of detecting the signature of
coalescing compact binaries in
the noise-dominated spectrum of a GW interferometer is the accurate modeling of
the expected signals. Only with an entire family of these theoretically
predicted template signals is it possible to filter large amounts of data
taken from the interferometers. In a ``matched-filter'' search (see
e.g.
\cite{Helstrom}), these data are convolved with the model signals and if the
agreement exceeds some predefined threshold one claims detection and
further
exploits theoretical predictions to estimate physical parameters of
the binary
system, such as component masses and spins.

In the case of a binary black hole (BBH) with comparable masses, at
least two different approaches are needed to describe
the full motion and radiated GW content from the system.
Post-Newtonian (PN) theory is an asymptotic weak field approximation that
treats black holes as point particles with a relative velocity $v$
that is small with respect to the speed of light $c$ (for details, see 
e.g.,~\cite{lrr-2006-4} and reference therein). The standard PN formulation is based
on expanding the relevant quantities (such as energy and GW flux) in terms of
the small parameter $v/c$. Depending on the details of the expansion,
resummation and integration of the resulting differential equations,
different
waveform models for the early inspiral are known, commonly denoted by TaylorT$n$
(with $n = 1, \ldots, 4$)
\cite{lrr-2006-4,Cutler:1992tc,Damour:1997ub,Buonanno:2005xu,Boyle:2007ft,
Buonanno:2009zt}, TaylorF2 \cite{Damour:2000zb, Damour:2002kr,
PhysRevD.72.029901, Arun:2004hn} and TaylorEt \cite{Gopakumar:2007jz,Hannam:2007wf}. 
A further inspiral waveform family is obtained by mapping the two body problem to an
effective one body (EOB) system with the appropriate potential
\cite{Buonanno:1998gg,Buonanno:2000ef,Damour:1997ub,Damour:2000we}.

All these analytical approximations break down in the strong gravity regime, and one
has to perform computationally expensive numerical calculations
in full general relativity to describe the complete dynamics.
Since 2005~\cite{Pretorius:2005gq,Campanelli:2005dd,Baker:2005vv} 
stable numerical-relativity (NR) simulations of a few orbits plus merger and 
ringdown to a final Kerr black hole have become a standard tool to 
consistently predict the last stages of a BBH coalescence~\cite{Hannam:2009hh}.
The exploration of the whole parameter space, however, has just begun and, for
instance, long simulations of systems with mass-ratios higher than $q = m_2/m_1 \sim 5$ 
are still exceptionally time-consuming.
For current overviews of the field see
\cite{Hannam:2009rd,Hinder:2010vn,Centrella:2010zf,McWilliams:2010iq}. 

An obvious goal is to combine PN and NR results to produce ``complete''
waveform models. Such signals contain physical information up to
frequencies higher than the pure PN templates, which becomes
increasingly important when the total mass of the system
increases. According to recent studies, binary neutron
stars as well as mixed black hole/neutron star binaries 
can be detected well by point-particle PN templates
\cite{Hinderer:2009ca,Pannarale:2011pk} assuming the current and
anticipated performance of the Laser Interferometer
Gravitational-wave Observatory (LIGO) \cite{Abbott:2007kv}. We
therefore focus on complete waveform models for BBH coalescences in
this paper. By including systems with small total masses in our
analysis, however, we
effectively consider the detection problem for a broad range of possible compact binary systems, 
although the extraction of all physical effects requires further modeling in the neutron star case.

Several approaches have already been suggested
to analytically build template families that include all the stages that the BBH
undergoes. The EOB family has been refined by adding extra parameters that
cannot be determined by PN calculations but are fixed by calibrating them to
highly accurate NR data. This combination of EOB and NR information
yields a description of the entire coalescence process in the time domain
\cite{Buonanno:2007pf,Buonanno:2009qa,Damour:2007vq,Damour:2008te,Damour:2009kr,
Pan:2009wj,Yunes:2009ef,Pan:2011gk}, often referred to as EOBNR. 
A different time-domain description based on standard
PN expansions was presented in \cite{Sturani:2010yv} as a step towards modeling
generic spin configurations. In this paper, we consider the direct
matching
of standard PN waveform models to NR data. In this approach, PN data are 
used up to some point in time or frequency and NR results are then
taken to describe the remaining part of the waveform. A phenomenological fitting 
of these ``hybrid'' waveforms can then be performed to obtain an analytical
closed formula (in the frequency domain) which interpolates between the physical
parameters of the 
hybrids~\cite{Ajith:2007qp,Ajith:2007kx,Ajith:2009bn,Santamaria:2010yb}.

All these procedures are subject to ambiguities and errors that limit the 
applicability of the final waveforms. 
Here we will focus on the error due to the PN contributions to the
hybrids only. Previous work has shown that the uncertainties in the NR
waveforms and in the hybridization procedure make a negligible
contribution to the overall
hybrid waveform error
budget~\cite{Santamaria:2010yb,Hannam:2010ky,MacDonald:2011ne}, and
as such we will estimate modeling errors on
the basis of the dominating inspiral part of the waveform.
In the absence of a well-defined notion of the PN error, however, we
have to account for it simply by considering two different PN
descriptions of the inspiral signal, which are equivalent to all
\emph{known} orders of their Taylor expansion. We then quantify the
effect of this ambiguous part of the waveform by calculating an
appropriately defined inner product (``match'') between both choices.
This data analysis-motivated measure leads directly to conclusions
about how useful hybrid waveforms are in the presence of an ambiguous
PN part, or conversely, what requirements have to be posed in order
to model waveforms accurately enough.

As an important application of this procedure
we started in a previous paper~\cite{Hannam:2010ky} addressing the
question of how long
numerical waveforms have to be in order to fulfill the accuracy
requirements for a PN/NR hybridization. Our analysis of
nonspinning binaries with mass-ratio $q \in [1,4]$ and
equal-mass binaries with spins (anti-)aligned to the orbital angular
momentum (with $\chi_i = S_i/m_2^2 \leq 0.5$) lead to the conclusion that 
NR simulations of such systems should cover 5 to 10 orbits to be used in 
hybrids that satisfy the minimal accuracy requirement for signal detection.
For larger mass-ratios and larger spins, our results suggested that far
longer numerical waveforms were required. 

However, that study was limited due to the following restriction:
The efficacy of a model in a search is determined by the best match between
the true waveform and {\it any waveform} in the search model. 
This best match (called the ``fitting factor'') should be calculated
not only by comparing two candidates but by
maximizing the match over all of the physical parameters of the model. 
With access to hybrids from discrete points in the parameter space,
we were only able to maximize the match over the total mass of the binary, 
and so our results were a (possibly very) conservative estimate of waveform
length requirements. 

Even stronger requirements were presented in a number of other
studies, where {\it no} maximization was performed at all (except over the initial phase
and time-of-arrival of the signal), with the intention of
determining the waveform length requirements not just for detection, but also for
parameter estimation. With these more stringent requirements, 
MacDonald \emph{et al.} \cite{MacDonald:2011ne} 
as well as Boyle~\cite{Boyle:2011dy}, concluded that NR waveforms generally 
have to be \emph{much longer} than currently possible to produce hybrids 
sufficiently accurate for both detection and parameter estimation. 
In addition, Damour \emph{et al.} \cite{Damour:2010zb}
presented a detailed comparison of phenomenological waveform
models \cite{Ajith:2009bn,Santamaria:2010yb} and a recent member
of the EOBNR
family~\cite{Damour:2009kr}. As part of their approach they find that
in particular systems with higher mass-ratio ($q \gtrsim 10$) can
 be combined accurately with a standard PN approximant in
the frequency domain only if the NR waveform contains {\it hundreds} of orbits.

In this paper, we study hybrid accuracy and NR waveform length
requirements
in the context of {\it fully optimized mismatches}, i.e., fitting
factors. 
Put differently, instead of quantifying the reliability of a
single waveform with fixed physical parameters we ask how accurate
the induced waveform family is at that point in the parameter space.
To do this, we first simplify the nonoptimized match calculation,
showing that it can be performed \emph{without} full numerical
waveforms. We then generalize our
procedure to optimize the match with respect to physical
parameters, and to then calculate the fitting factor that is necessary
to make estimates of NR waveform length requirements that are
meaningful for GW searches. 
By looking at the parameter bias 
between the best-match waveform and the target signal, we also 
gain some insight into the parameter estimation errors due to the 
uncertainties in the waveform modeling process.

In the following sections, we will develop this procedure step by
step, starting with the stringent assumptions previous results were
based on and subsequently relaxing them until we reach the final
result.
After providing a mathematical definition of the (mis)match as our
notion of error (Sec.~\ref{sec:prelim}), 
we show in Sec.~\ref{sec:hybMM}  that
the mismatch between two hybrids is determined by the PN uncertainty
and the relative power between the NR and PN parts of the signals.
Thus, our accuracy estimate requires only amplitude information in the
NR regime and we derive how this can be incorporated, including the
effect of possible time and phase shifts of the entire waveforms. 
(This is similar to the procedure developed by
Boyle in~\cite{Boyle:2011dy}, where instead of NR data, EOBNR
signals are taken as ``ersatz'' waveform. We use a slightly more
general approach by incorporating only amplitude information of
the phenomenological model~\cite{Santamaria:2010yb}.) 
Along the way we compare with previous results in the literature, showing that we 
fully agree on nonoptimized mismatch errors. 

When we finally optimize 
these mismatches with respect to physical parameters in Sec.~\ref{sec:FF}, 
we find that the corresponding errors for waveform families are much smaller 
than assumed so far. In particular, based on our estimates (that are
mainly limited by the choice of PN families compared to each other) we
conclude that NR simulations that cover $\sim 10$ orbits are
probably acceptable for most astrophysical applications during the 
Advanced detector era. 
This includes nonspinning binaries (for which significant improvements in 
PN approximants are less likely in the next five years), where we explicitly show that this relatively small number of 
NR orbits is sufficient up to at least $q =10$, and with astrophysically reasonable restrictions even for $q=20$ and above. 
We also adopt the view that, since typical 
simulations will be of comparable lengths over the next five years, 
our focus here and in future work should not be on prescribing ideal (and unrealistic) 
waveform lengths, but on determining the limits of the science that we
can do with them.

\section{Preliminary considerations} \label{sec:prelim}

We shall address the question of accuracy of BBH hybrid waveforms in
the following sense. Let
\beq
  h = h_+ - i \, h_\times
\eeq
be the complex GW strain that combines the plus and cross polarization
of the GW
as the real and imaginary part, respectively. It is constructed from
its PN
description $h_{\rm PN}$ and the NR part $h_{\rm NR}$. We assume that
the
transition from $h_{\rm PN}$ to $h_{\rm NR}$ is enforced at a single
frequency
\beq
 \tilde h(f) = \left\{ \begin{array}{ll}
                     \tilde h_{\rm PN}(f) ~, & \textrm{for}~ f\leq f_m
\\[4pt]
		    \tilde h_{\rm NR}(f) ~, & \textrm{for}~ f > f_m
                    \end{array}
\right.~, \label{eq:transition}
\eeq
where $\tilde h$ denotes the Fourier transform of $h$ and $f_m$ is the
matching
frequency. Such a procedure can be employed in a direct Fourier-domain
construction of the hybrid \cite{Santamaria:2010yb}, but it is also
approximately true for time-domain hybrid constructions. In the latter
case,
the transition is carried out at a time $t_m$, where the instantaneous
frequency is
$\omega(t_m) = \frac{d \arg h}{dt}  = 2 \pi f_m$. Then, for
(\ref{eq:transition}) to be true, we have to assume that
\begin{enumerate}
 \item the transition frequency in the Fourier domain is equal to the
instantaneous matching frequency calculated in the time domain;
 \item the signal at times $t<t_m$ only significantly affects the
Fourier domain for $f < f_m$ and $t > t_m$ correspondingly determines
the wave
for $f>f_m$.
\end{enumerate}
These assumptions are not trivial since the Fourier integral is a
``global'' transformation. However, it was shown that assuming such
a stationarity is reasonable in a regime where both PN and NR are
valid
\cite{Santamaria:2010yb} and time- and frequency-domain construction
methods lead to very similar results \cite{Hannam:2010ky}.
 
The final hybrid waveform is subject to several errors, and we account
for these
errors here simply by the fact that one could have taken slightly
different
ingredients $h_{\rm PN}$ and $h_{\rm NR}$ for the same physical
scenario. These
could be different post-Newtonian approximants and numerical data from
different
codes or different resolutions. Denoting the different waveform models
by $h_1$
and $h_2$, we calculate the mismatch
\begin{align}
 \mathcal M &= 1 - \mathcal O (h_1,h_2) = 1- \frac{\langle h_1, h_2
\rangle}{\|
h_1 \| \, \| h_2 \|
}\\
 &= 1 - \max_{\phi_0, t_0}\left[ 4 \operatorname{Re} \int_{f_1}^{f_2}
\frac{\tilde
h_1(f) \, \tilde h_2^\ast (f)}{S_n(f)} \, 
\frac{d\!f}{\| h_1 \| \| h_2 \|} \right], \label{eq:mm_int}
\end{align}
where $\phi_0$ and $t_0$ are relative phase and time shifts between
the
waveforms and $\| h \|^2 = \langle h,h \rangle$. $S_n$ is the noise
spectral
density of the assumed detector, $^\ast$ indicates the complex
conjugation and
$(f_1, f_2)$ is a suitable integration range. $\mathcal O$ is called
the
overlap (or match) of the two waveforms. Throughout this paper, we
will follow
the choices of our preceding work \cite{Hannam:2010ky}, i.e., $f_1 =
20\textrm{Hz}$ and $S_n$ is given by the analytic fit of the design sensitivity of
Advanced LIGO \cite{Ajith:2007kx}. 
The upper integration bound $f_2$ is
given by our waveform model, and we use $f_2 = 0.15/M$, although the
results do not depend sensitively on this value ($M$ is the total mass
of the binary).

Broadly speaking, the mismatch indicates how ``close'' $h_1$ and
$h_2$ are. Smaller values for $\mathcal M$ represent
smaller errors in the waveform model, given that $h_1$ and $h_2$ are
approximations of the same signal. 
Direct conclusions can be drawn from calculating the mismatch: If
$\mathcal M$ is less than some threshold, we regard the final hybrid
as
\emph{accurate enough} for the purpose in question. For a maximum loss
of 10\%
of the signals in the detection process, we can accept a mismatch of
$\approx 3\%$,
disregarding the addition from
a discrete template spacing. If we account for the latter, one may
decrease
the accepted mismatch in the waveform modeling to 1.5\% (see a similar
discussion in \cite{Hannam:2010ky})
or even 0.5\% as suggested in \cite{Lindblom:2008cm}. 

A generally more
stringent
requirement is that the uncertainty we have in the modeling is
\emph{indistinguishable} by the detector. Such a statement is
obviously
dependent
on how ``loud'' the signal is in the detector. As discussed in
\cite{Lindblom:2008cm} and further detailed in
\cite{Damour:2010zb,Lindblom:2009ux} we
can write
the indistinguishability criterion as
\beq
  \| h_1 - h_2 \|^2 < \epsilon^2~, \label{eq:indist_h}
\eeq
where the waveforms are optimally aligned in the sense of
(\ref{eq:mm_int}) and $\epsilon$ parametrizes the effective
noise-increase due to model uncertainties. The minimal requirement for
$h_1$ and $h_2$ to be indistinguishable is $\epsilon =1$, although
\cite{Damour:2010zb} argues that $\epsilon \sim 1/2$ and probably less
are more reasonable thresholds. 
Manipulating (\ref{eq:indist_h}) under the assumption of equal norms
leads to the
equivalent inequality (see the calculation in
\cite{McWilliams:2010eq}) 
\beq
  \mathcal M < \frac{1}{2 \rho_{\rm eff}^2}~, \label{eq:indist_M}
\eeq
where $\rho_{\rm eff} = \| h \|/\epsilon$ is the effective
signal-to-noise ratio (SNR) of
the signal.

When we later calculate $\mathcal M$ as a measure of the error
in hybrid waveforms, we can set various thresholds based on $\mathcal
M <\mathcal M_{\rm max}$ or Eq.~(\ref{eq:indist_M}) to evaluate the
reliability of current models.
A potentially very useful application is then to conclude which
matching frequency is needed
(i.e., how
long do the numerical waveforms have to be) to ensure the desired
accuracy.

\section{Hybrid mismatches }
\label{sec:hybMM}

Having introduced the mismatch between supposedly equivalent waveform
models as our notion of error, we shall devote this section to
simplifying the mismatch calculation of two hybrid waveforms,
optimized only with respect to a relative time and phase shift. As we
have pointed out in the introduction, this is not the complete
procedure to assess the model accuracy in terms of signal detection
because the optimization with respect to physical parameters is
not considered. However, we first need to develop some insights into
this
simpler procedure to eventually generalize it in the next section. The
results presented in Sec.~\ref{sec:hybMM_app} therefore have mainly
illustrative character, showing that our simplified approach fully
agrees with previously published results, but it leads to overly
conservative requirements, e.g., for the length of NR waveforms.

\subsection{General procedure} \label{sec:hybMM_general}

Before we calculate mismatches for many different scenarios, we
establish a few more assumptions to gain some
insights on the structure of Eq.~(\ref{eq:mm_int}). These will allow
us to 
propose an approximation to the mismatch between two hybrid
waveforms that can be calculated {\it without} the need for 
any NR data.  

In addition to (\ref{eq:transition}), we further assume:%
\begin{enumerate}
\setcounter{enumi}{2}
 \item following \cite{Santamaria:2010yb,Hannam:2009hh} we regard the
error on
the NR side as small, negligible compared to the uncertainties PN
introduces up
to currently practical matching frequencies.
 \item independent of the PN approximant that is used, the norm of the
waveforms
are to high accuracy the same (i.e., only the phase is affected).
    This is reasonable to take as a good approximation, because the
amplitude
description in PN is usually formulated as a function of the orbital
frequency
\cite{Berti:2007nw,Blanchet:2008je,Arun:2008kb}
(which we again identify with the content on the Fourier side as well)
and the
mismatch is much more sensitive to phase differences than to amplitude
discrepancies.
\end{enumerate}

Let us now consider a BBH system with fixed physical parameters.
Our error measurement assumes the construction of two
hybrid waveforms that differ in the PN part only. Their overlap reads
\begin{align}
 &\mathcal O(h_1, h_2) = \max_{\phi_0, t_0}\left[ 4 \operatorname{Re}
\int_{f_1}^{f_2}
\frac{\tilde
h_1(f) \, \tilde h_2^\ast (f)}{S_n(f)} \, 
\frac{d\!f}{\| h_1 \| \| h_2 \|} \right]  \label{eq:Ov}\\
& = \max_{\phi_0, t_0}\left[ 4 \operatorname{Re} \int_{f_1}^{f_2}
\frac{\vert A_1 A_2 \vert}{S_n} e^{i(\phi_1 - \phi_2)}  \, e^{i(2 \pi
f \, t_0 + \phi_0)} 
\frac{d\!f}{\| h_1 \| \| h_2 \|} \right]~, \nonumber
\end{align}
where $A_i = \vert \tilde h_i \vert$ and $\phi_i = \operatorname{arg} \tilde
h_i$. The effect of a time and phase shift of one waveform with
respect to the other is explicitly written out in the second
exponential term. 

Assuming two PN models (PN1 and PN2) combined with the same NR
waveform we trivially obtain the phase difference
\beq
 \phi_1 - \phi_2 = \left\{ \begin{array}{cl}
                           \phi_{\rm PN1} - \phi_{\rm PN2} &~, f < f_m \\
			0  &~, f \geq f_m~.
                          \end{array} \right. \label{eq:phasediff}
\eeq
Note that (\ref{eq:phasediff}) is only true for one particular
alignment of the two waveforms, any other relative shift in time or
phase
leads to an additional dephasing, also beyond $f_m$. Since we
have separated this effect explicitly in (\ref{eq:Ov}), we are,
however, free to write $\phi_1 - \phi_2$ as in (\ref{eq:phasediff}).
The open question is the functional form of the PN phase-difference
(or simply the PN phase error) in the case
where the NR part of $h_1$ and $h_2$ are perfectly
aligned. Here we have to apply an actual matching procedure, although
we can use
any preferred method \emph{without} having NR data at
hand. The key property of (\ref{eq:phasediff}) we are exploiting is
that only PN-PN differences are taken into account, and a direct PN-NR
comparison is not necessary. The only input we need from NR
simulations is the amplitude $\vert \tilde h \vert = A_1 = A_2$ for $f
> f_m$. A good estimate for that can be taken from phenomenological
models, such as \cite{Ajith:2009bn} or \cite{Santamaria:2010yb}, where
the Fourier-domain amplitude is approximated by a closed-form
analytic description. A similar
approach was recently suggested by Boyle \cite{Boyle:2011dy} who
realized that it is sufficient to combine PN approximants with
\emph{ersatz} NR data which he takes from the EOBNR model
\cite{Buonanno:1998gg,Buonanno:2000ef,Buonanno:2009qa,Pan:2009wj}.
We independently derive an algorithm here that is based on the same
perceptions but highlights that no NR phase information \emph{at all}
is needed. 

The final global time and phase shift used in (\ref{eq:Ov}) to
maximize the overlap is simply a (phase shifted) inverse Fourier
transform of the remaining integrand. Its maximal real part is
obtained by choosing $\phi_0$ (for any $t_0$) such that the generally
complex number lies on the real axis.

Based on that, our final algorithm for estimating hybrid mismatch errors caused
by the uncertainty in the PN model is the following
\begin{enumerate}
 \item Calculate the two different PN waveforms expressing the uncertainty to be
quantified.
 \item Apply the matching procedure such that one PN approximant is
matched to the other at $f_m$ (as if it were the NR waveform).
 \item Fourier transform the aligned PN waveforms and keep the data for $f
\in [f_1 , f_m]$.
 \item Complete the waveforms in the Fourier domain by using an
existing
expression for the amplitude in the range $f \geq f_m$, e.g. from
\cite{Santamaria:2010yb,Ajith:2009bn} or from a short NR simulation. Set the
phase in this regime to $0$ (or any other function, but equal for both
$\tilde h_1$ and $\tilde h_2$).
 \item Calculate the overlap of $\tilde h_1$ and $\tilde h_2$ by
maximizing the magnitude of the inverse Fourier transformation.
\end{enumerate}
 
\begin{figure}
 \includegraphics[width=0.87\columnwidth]{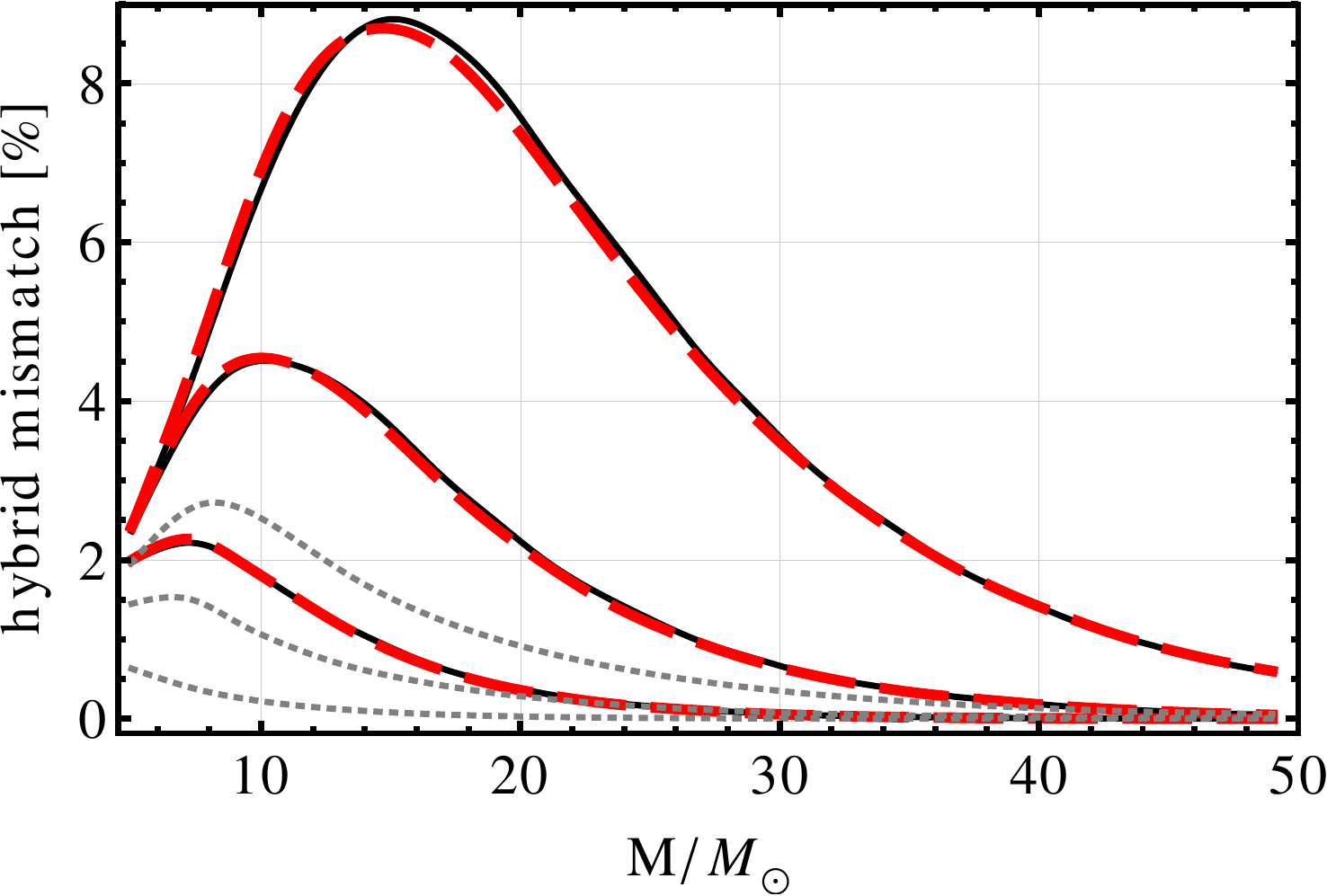}
\caption{Hybrid mismatches in the equal-mass, nonspinning case. Black solid
lines are mismatches of actual TaylorT1/T4+NR hybrids, whereas the
dashed (red online) lines are our estimates obtained without directly using
 any NR data (NR amplitude taken
from a phenomenological model). The gray dotted lines describe the PN
mismatch contribution derived in Eq.~(\ref{eq:mismatch_simple}) that
does not include a possible dephasing of the NR part. 
The matching frequencies for each set are from bottom
to top $M \omega_m =
0.04,0.06,0.08$.}
\label{fig:hybEst}
\end{figure}

To test the efficacy of our approach, we compare our estimate with
the mismatch of actual hybrids consisting of either the TaylorT1 or
TaylorT4 approximant (in the form detailed in
\cite{Santamaria:2010yb}; see also Sec.~\ref{sec:hybMM_app}) and the
numerical data from the
\texttt{SpEC} equal-mass run \cite{Scheel:2008rj,SpEC:wfs}. The
matching frequencies are chosen as $M \omega_m = 2 \pi M f_m \in
\{0.04,0.06,0.08\} $, and the stitching procedure is carried out in
the Fourier domain as explained in \cite{Santamaria:2010yb}. The
agreement illustrated in Fig.~\ref{fig:hybEst} is excellent in all
cases.
As expected by the relatively small effect of the amplitude on the
mismatch calculation, our method proves to be fairly robust with
respect to the
chosen amplitude description in the NR regime. In fact, the dashed
lines in Fig.~\ref{fig:hybEst} use the phenomenological model detailed by
Santamar\'ia {\it et al.} in
\cite{Santamaria:2010yb} but there is no noticeable difference when
we use the model presented by Ajith {\it et al.} in
\cite{Ajith:2009bn}. %

\subsection{Mismatch contributions}

The method presented above can readily be applied to estimate the
uncertainty of hybrids with the caveats mentioned at the beginning
of Sec.~\ref{sec:hybMM}, and we shall do so in
Sec.~\ref{sec:hybMM_app}. For now, however, let us 
manipulate the mismatch (\ref{eq:mm_int}) further to separate the
various contributions to it. We make this important aside to point
out that, although only the PN contribution is considered as
ambiguous in our approach, its influence on the final waveform
error is twofold: directly through the (power-weighted) PN mismatch
and in terms of an additional dephasing, also of the ``exact''
high-frequency part.

We can see these two effects separately through the following 
instructive lower bound on $\mathcal M$ which is obtained
under the assumptions detailed above.
\begin{align}
 \mathcal M  & = 1 -  \frac{4}{\| h \|^2} \max_{\phi_0,
t_0}  \left[  \operatorname{Re} \int_{f_1}^{f_m} \frac{\tilde h_1 \,
\tilde
h_2^\ast}{S_n}d\!f  + \int_{f_m}^{f_2} \frac{\vert \tilde  h
\vert^2
}{S_n} d\!f \right]\nonumber \\
& \geq 1 -  \frac{\langle h_1, h_2 \rangle }{\| h \|^2_{(f_1,f_m)}}
\frac{\|
h \|^2_{(f_1,f_m)}}{\| h \|^2} - \frac{\| h \|^2_{(f_m,f_2)}}{\| h
\|^2}\nonumber \\
 & =   \frac{\| h \|^2_{(f_1,f_m)}}{\| h \|^2} \, \mathcal M_{\rm
PN}~.
\label{eq:mismatch_simple}
\end{align}
Here we introduced the notation $\| h \|^2_{(a,b)}$ to specify the
integration
range. $\mathcal M_{\rm PN}$ is the mismatch of the PN part only,
restricted to
$f < f_m$. In the first line of (\ref{eq:mismatch_simple}) we use the
fact that
the amplitudes agree (in fact, we do not require pointwise agreement,
only the
norm is assumed to be the same) and that $h_1 = h_2$ for $f > f_m$.
The second
line is a lower estimate because the maximization was
originally
carried out by shifting the entire waveforms relative to each other,
whereas now
we allow the maximization over the PN part alone. The final step
involves the obvious
relation $\| h \|^2 = \| h \|^2_{(f_1,f_2)} = \| h \|^2_{(f_1,f_m)} +
\| h
\|^2_{(f_m,f_2)}$.

The interpretation of (\ref{eq:mismatch_simple}) is straightforward:
the
mismatch of hybrids is
determined by the uncertainty of PN $[$restricted to the 
frequency range $(f_1, f_m$)$]$ multiplied by the fraction of power
that is coming from the PN part of the wave signal.
This fundamental error, independent of the actual PN/NR fitting, is 
directly inherited from the differences of standard PN approximants
and any PN/NR matching cannot be better than the result of
(\ref{eq:mismatch_simple}).
Therefore, one might think that analyzing the overlaps or fitting
factors (or whatever strategy is appropriate) of different
post-Newtonian
approximants directly leads to conclusions of how reliable
the hybrid
is for a particular choice of $f_m$. 
When we compare, however, the
mismatch of actual hybrid waveforms with the estimate
(\ref{eq:mismatch_simple}) we find that the latter is considerably
less than $\mathcal M$. An illustration of that is included in
Fig.~\ref{fig:hybEst}, where we show the lower bound
(\ref{eq:mismatch_simple}) in comparison with the actual (and
accurately estimated) mismatches. 

Why is the hybrid disagreement that much greater than what is expected
from PN
in the given frequency range? The reason can be identified from the
derivation of (\ref{eq:mismatch_simple}), where we effectively allow
an optimal alignment (for each $M$) of both PN models while
\emph{independently} keeping the NR part \emph{perfectly
aligned}. In a true hybrid mismatch calculation, one the other hand, a
time and/or phase shift always affects the \emph{entire PN+NR hybrid},
and an optimal alignment of one part leads to a dephasing of the
other. This effect is not caused by an erroneous matching,
but an illustration of the fact that the optimal choice of
$t_0$ and $\phi_0$ in the sense of Eq.~(\ref{eq:mm_int}) is
mass (frequency)-dependent for the PN models we consider.

Finally, by considering the obvious generalization of
(\ref{eq:mismatch_simple}),
\beq
 \mathcal M \geq  \frac{\| h \|^2_{(f_1,f_m)}}{\| h \|^2} \, \mathcal
M_{\rm
PN} + \frac{\| h \|^2_{(f_m,f_2)}}{\| h \|^2} \, \mathcal M_{\rm NR}~,
\label{eq:hybMM_2parts}
\eeq
we can identify the three main contributions to the hybrid
uncertainty: The PN and NR error, each weighted by the power
they contribute to the signal and the misalignment caused by the fact
that in the hybridization procedure the PN wave is aligned at high
frequency which is potentially different from the optimal alignment
for lower frequencies. The procedure introduced in
Sec.~\ref{sec:hybMM_general} automatically takes the dominant PN
error and possible misalignments (also of the NR part) into account.

\subsection{Application} \label{sec:hybMM_app}

Now that we have established an algorithm to predict the full
waveform mismatches, we can exploit the computationally cheap
procedure and calculate $\mathcal{M}$ for many
different physical scenarios. Our aim is to show how ``reliable''
the final combination of PN and NR waveforms is in different points
of the parameter space, assuming that the physical parameters
are fixed from the outset.

First, let us highlight again that ideally, we
are interested in the mismatch of the approximate waveform model to
the true one. Since we cannot calculate the
latter (which would also make the whole discussion pointless), we
estimate the PN uncertainty by calculating the mismatch between
different approximants.
This can certainly be no more than a rough estimate since we are not
aware of any principle that would guide us to which approximants at
which PN order should be compared in order to obtain a well-defined
notion of the PN error.

To still reach some understanding of the uncertainty in currently used
high-order PN models we present the anticipated hybrid
mismatches when approximants commonly denoted by
TaylorT1, TaylorT4 and TaylorF2 are used. TaylorT1 and T4 are
solutions of ordinary differential equations in the time domain describing
the adiabatic inspiral of the BBH on quasicircular orbits, whereas
TaylorF2 is a frequency-domain description based on the stationary
phase approximation. Details on these approximants can be found, e.g.,
in \cite{Buonanno:2009zt,Boyle:2007ft} and references therein. We
mainly employ the equations presented in \cite{Santamaria:2010yb}, but
with an updated 2PN spin-spin contribution from \cite{Arun:2008kb},
see \cite{Brown:2007jx} for a collection of explicit expressions.
Throughout this paper, we always employ the highest currently
determined PN order, i.e., 3.5PN accurate phasing with spin
contributions up to 2.5PN (and incomplete terms at higher order) and
the 3PN amplitude expansion
\cite{Blanchet:2008je} including up to 2PN spinning corrections
\cite{Arun:2008kb}.

As in the construction of phenomenological models, we restrict the
parameter space to black holes with comparable masses and spins
aligned or antialigned with the orbital angular momentum of the
binary $\bm L$ (with its unit vector denoted by $\bm {\hat L}$). 
Then, each spin can be parameterized by just one dimensionless
quantity,
\beq
 \chi_i = \frac{\bm{S_i} \cdot \bm{\hat L}}{m_i^2}~, \quad i=1,2,
\eeq
where $m_i$ and $\bm S_i$ are mass and spin of the individual black
hole, respectively. By exploiting a degeneracy in the spins,
as observed in \cite{Vaishnav:2007nm,Reisswig:2009vc}, the
parameter space can be further reduced, and we only use the
mass-weighted total spin
\beq 
\chi = \chi_1 \, m_1/M  + \chi_2 \, m_2/M \label{eq:chi_definition}
\eeq 
and the symmetric mass-ratio
\beq
 \eta = \frac{m_1 m_2}{M^2} \label{eq:symm_massratio}
\eeq
to label the different physical setups. (In fact, in the following
analyses, each point with fixed $\chi$ is represented by $\chi_1
= \chi_2 = \chi$.)

\begin{figure*}
 \includegraphics[width=0.4\textwidth]{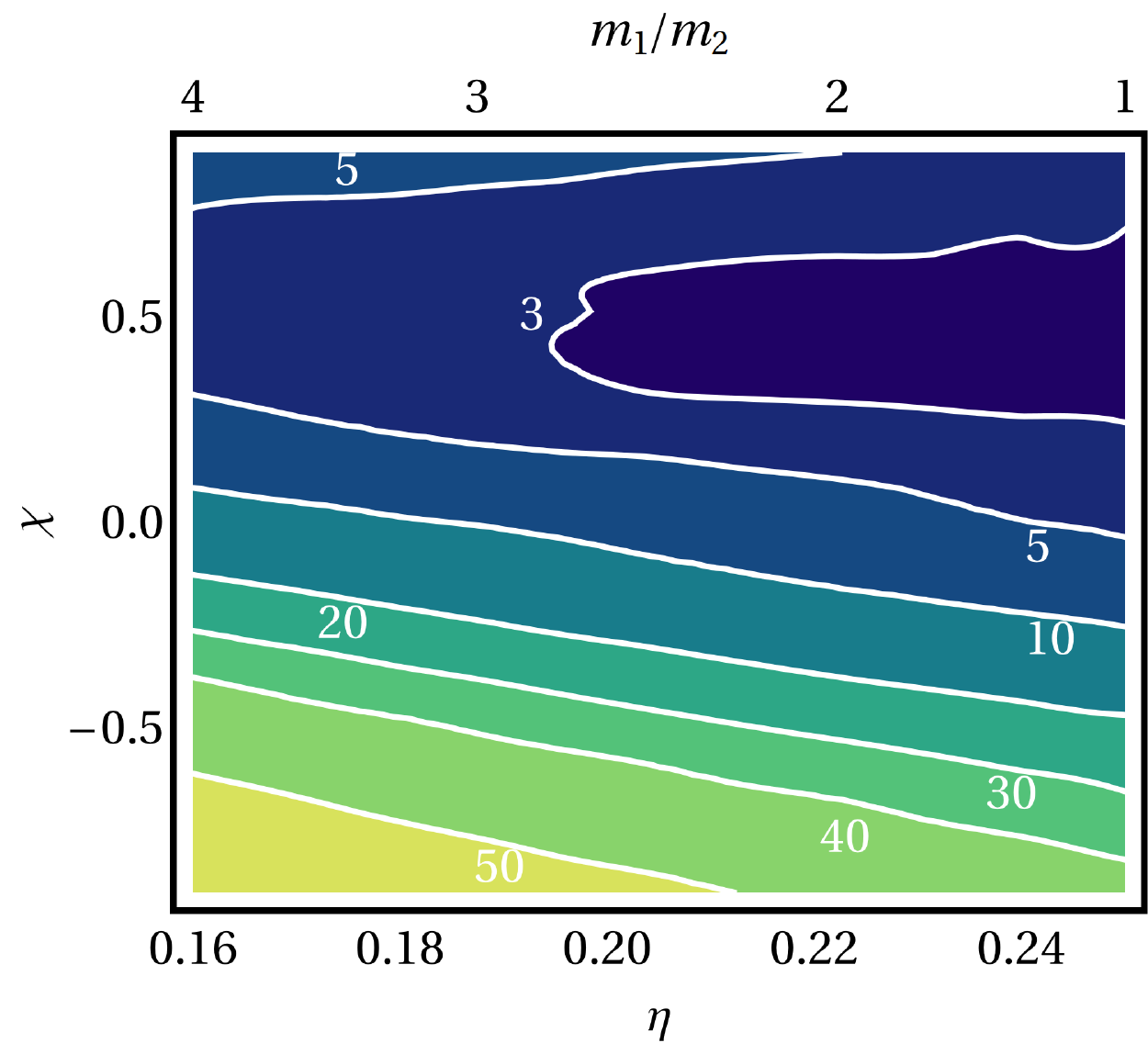} \hfil
\includegraphics[width=0.4\textwidth]{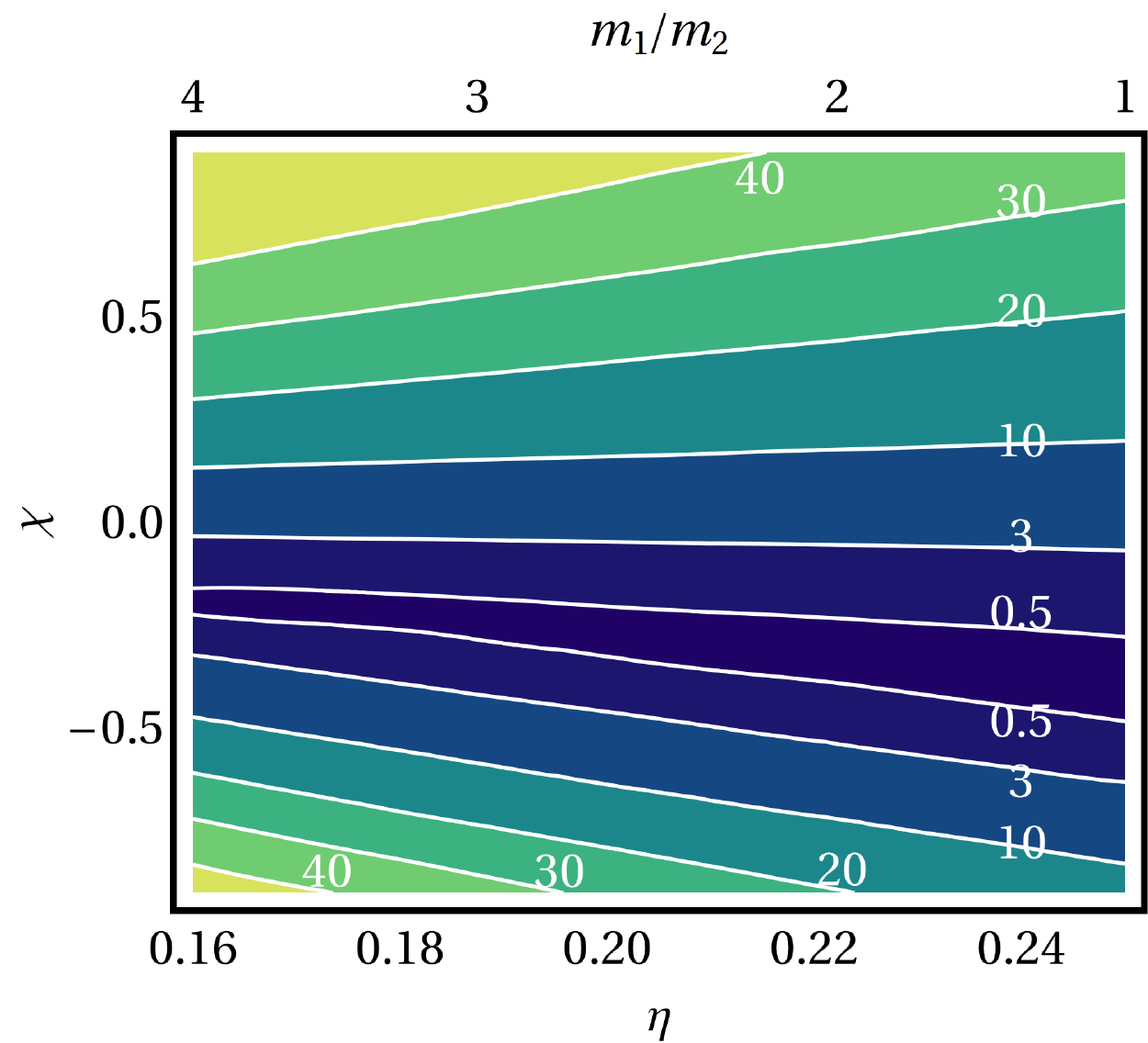}
\caption{Contour plots of the mismatch (in \%) between different
fictitious
hybrids, as a function of the symmetric mass-ratio $\eta$ and equal
aligned spins with dimensionless magnitude $\chi$. \emph{Left panel:}
PN part either defined by the TaylorT1 or TaylorT4 approximant.
\emph{Right panel:} Comparison of TaylorT1 and TaylorF2 in the PN
part.}
\label{fig:MMContours}
\end{figure*}

To assess how the accuracy of currently feasible hybrid waveforms
varies in the parameter space, we apply the algorithm
outlined in Sec.~\ref{sec:hybMM_general} for different mass-ratios
ranging from equal masses to 4:1, with spin magnitudes from
$-0.9$ to $0.9$ in each case. For every pair $(\eta, \chi)$ one
obtains mass-dependent mismatches in the form of
Fig.~\ref{fig:hybEst} that generally increase with increasing
matching frequency $M \omega_m$.

Several plots illustrating this behavior can already be found in the
literature. Contour plots of the mismatch as a function of mass and
matching frequency are the main result of Boyle~\cite{Boyle:2011dy},
and we
obtain similar results by continuously varying $M \omega_m$, e.g., in
Fig.~\ref{fig:hybEst}. Taking the maximum mismatch with respect to the
total mass instead (i.e., only considering the peaks in
Fig.~\ref{fig:hybEst}), Fig. 4 by Damour, Nagar and Trias in
\cite{Damour:2010zb} shows the
inaccuracy of TaylorF2 hybrids compared to EOBNR as a
function of the matching frequency. Fig.~11 by
MacDonald, Nissanke and Pfeiffer in \cite{MacDonald:2011ne}
presents a similar study with Taylor approximants and actual NR data.
Given some slightly different choices in our approaches (especially
lower cutoff frequency and detector noise curve) the
results we obtain are fully consistent with the numbers presented in
the articles mentioned.  

Generally, the conclusions
\cite{Damour:2010zb,Boyle:2011dy,MacDonald:2011ne} draw are sobering
regarding GW detections and parameter estimation.
The mismatches found are too high, current numerical relativity
waveforms are \emph{by far} too short and hybrids are consequently too
inaccurate. 
In the following, we illustrate the basis of these statements and 
expand the existing knowledge by exploring the parameter space.
To reduce the dimensionality of the problem, we calculate the
maximum of the mismatch with respect to the total mass and fix $M
\omega_m =
0.06$ (which corresponds to 10 GW cycles before the maximum
of $\vert h(t) \vert$ in the equal-mass case). 

In Fig.~\ref{fig:MMContours} we show contour plots that compare 
either TaylorT1 with TaylorT4 hybrids or TaylorT1 with TaylorF2 hybrids.
The matching frequency is fixed at $M \omega_m = 0.06$. Certainly, we could include many more variants
of 
PN approximants (including different versions of EOBNR), but
we
find it sufficient to present some general conclusions that
become
already clear from the examples chosen here. As reported before
\cite{Hannam:2010ky,Damour:2010zb} we see that deviating from
equal-mass cases, the
disagreement generally becomes larger. This effect is even more
pronounced when increasing spin magnitudes are considered.
Heuristically we can understand the worse performance for increasing
spins by the simple fact that spin contributions are only included
up to 2.5PN order, whereas nonspinning terms are known up to
relative 3.5PN order. 
Surprisingly, the `island' or `band' of minimal mismatch does not
occur strictly around vanishing spin magnitudes, indicating that
different approximants can \emph{by chance} agree extremely well in
some portions of the parameter space. For completeness, let us report
that the TaylorT4/TaylorF2 mismatch yields a pattern similar to the
right panel of Fig.~\ref{fig:MMContours} but with minimal values
moved to weakly positive spins.

The conclusions suggested by Fig.~\ref{fig:MMContours} and
results from previous work
\cite{Santamaria:2010yb,MacDonald:2011ne,Boyle:2011dy} are 
indeed disappointing.  If the mismatches caused by different PN
approximants
actually represent a reasonable estimate for the uncertainty in
currently
practical hybrid waveforms, then values up to $\mathcal M
\approx 50\%$  are certainly unacceptable. 
Reducing the matching frequency, thereby demanding longer NR
waveforms, does reduce the mismatch everywhere, but it leads to
unrealistic requirements in many portions of the parameter space.

\begin{table}
\begin{ruledtabular}
 \begin{tabular}{|c|r|c|c|c|c|}
\multirow{2}{*}{~~$ \bm q$~~}  & \multirow{2}{*}{~~$\bm \chi$~~}  &
\multicolumn{2}{c|}{$\bm{M\omega_m [\times 10^{-2}]}$ } &
\multicolumn{2}{c|}{$\bm{ M_{\rm min}/M_{\bm \odot}}$ %
($M\omega_m = 0.06$)} \\ 
& & $\mathcal M < 3\%$ & $\rho_{\rm eff} < 20$ &
$\mathcal M < 3\%$ & $\rho_{\rm eff} < 20$ \\ \hline
 1:1 & 0.0 & 3.93 \emph{(23)} & 1.15 \emph{(212)} & 15 & 40 \\
\hline 
1:2 & 0.2 & 2.40 \emph{(68)} & 0.99 \emph{(313)} & 25 & 49 \\
\hline 
1:3 & 0.5 & 1.70 \emph{(155)} & 0.84 \emph{(499)} & 33 & 57 \\
\hline
1:4 & 0.8 & 1.38 \emph{(268)} & 0.75 \emph{(730)} & 38 & 61
 \end{tabular}
\end{ruledtabular}
\caption{Faithfulness of hybrid waveforms based on a
TaylorT1/TaylorF2 comparison for selected physical parameters.
The required matching frequency is reported if either a 3\%
maximal mismatch $\mathcal M$ can be tolerated or if the error should
be indistinguishable for SNRs less than 20, see
(\ref{eq:indist_M}). The parentheses indicate the number of GW
cycles to the maximum of $\vert h(t)\vert$. The two right columns
assume $M\omega_m = 0.06$ and give the minimal mass, where the
waveforms are accurate enough in the sense described above.}
\label{tab:mismatch_values}
\end{table}

To illustrate this, Table~\ref{tab:mismatch_values} addresses two
important questions by analyzing the TaylorT1/TaylorF2 hybrid
mismatches in selected points in the parameter space. First,
what is the
required matching frequency if a desired accuracy has to be fulfilled?
Note that due to our algorithm we overcome the restriction of
currently available NR waveform lengths that the authors in
\cite{Santamaria:2010yb,MacDonald:2011ne} were facing. We also do not
rely on assuming a particularly promising ``candidate waveform'' to
act as a long NR waveform as was done in
\cite{Hannam:2010ky,Boyle:2011dy}. In
fact, phase information above $M \omega_m$ is not required and does
not enter the result; we can simply apply our algorithm to
arbitrarily small matching
frequencies. For each set of parameters we maximize the mismatch with respect to the
total mass $M$ (which we, however, restrict to $M \geq 5 M_\odot$
for computational reasons) and thus obtain the 
monotonically increasing function $\max_M \mathcal M (M\omega_m)$. 
By demanding either $\mathcal M < 3\%$ as the most relaxed requirement
or the more stringent case of indistinguishable differences for
effective SNRs of at most 20 [see (\ref{eq:indist_M})] we obtain the
values given in Table~\ref{tab:mismatch_values}. In parentheses we
also give the
number of gravitational-wave cycles from $d\phi_{\rm GW}/dt = 
\omega_m$ to the maximum of $\vert h (t) \vert$ as predicted by the
phenomenological waveform model \cite{Santamaria:2010yb}. 

It is unlikely that the typical length of ``long'' numerical waveforms will change
by an order of magnitude before the advent of Advanced LIGO, and so a more
practical question is: given a currently achievable NR
waveform length, in which mass-range is the PN+NR hybrid accurate
enough? As an example we assume again a matching frequency of $M
\omega_m = 0.06$ and show on the right-hand side of
Table~\ref{tab:mismatch_values} the minimal masses the hybrid is
accurate for in the sense detailed above. For comparison, the pure NR
part occupies the entire frequency band down to 20Hz for masses $M
\geq 97 M_\odot$.
Note that, distinct
from \cite{Damour:2010zb}, we do not consider errors \emph{above} $M
\omega_m$ since we are concerned with \emph{hybrids} and not
possibly fitted closed-form waveform models that introduce
additional errors. Therefore, our values for $M_{\rm min}$ are
less than the corresponding results in \cite{Damour:2010zb} that are
based on the comparison of EOBNR and the phenomenological
model of 
\cite{Ajith:2009bn}.

The obvious message from Table~\ref{tab:mismatch_values} is that
\emph{in general} extremely long NR simulations would be needed to
overcome the intrinsic uncertainty in standard PN formulations for
given physical parameters. For NR waveforms containing so many cycles 
our assumption that their intrinsic error can be neglected is
possibly no longer valid, which would lead to even higher modeling errors. Anyway, the numbers presented are only an 
``order of magnitude'' estimate in this most
conservative approach. The reader should always keep in mind that our
notion of error is based on comparing different, at highest
available order consistent PN descriptions and especially concrete
statements for particular points in parameter space may be spoiled by
an (un)fortunate choice of approximants (see a similar
discussion in \cite{Hannam:2010ky}). More importantly, as we shall
show in the next section, fixing the physical parameters of the
waveforms from the outset greatly overestimates the uncertainty for
signal detection.

\section{Fitting factors} \label{sec:FF}

\subsection{General}

The accuracy assessment presented in Sec.~\ref{sec:hybMM}
only allows for very limited conclusions about the actual utility of
hybrid waveforms in various applications. Apart from the restrictions
coming from our limited understanding of the PN error there is also
an important fact we have neglected so far: in astrophysically
relevant applications the knowledge of physical parameters like
total mass, mass ratio and spin is never \emph{exact}. If a set of
hybrid waveforms constitutes a \emph{waveform family} which is used
to extract information from an unknown signal, then the standard
matched-filter procedures rely on varying (and maximizing with respect
to) such parameters. The accuracy of the predicted ``best-fit''
parameters is once again limited by the detector noise and the
modeling error and even if the latter exceeds the first, one may
still argue that a tolerated bias does not significantly reduce the
scientific output from GW detections.

In this section we shall therefore consider combinations of NR
data with a particular PN approximant as the ingredients of an entire
manifold of waveforms, parametrized by an absolute time and phase
scale ($t_0$ and $\phi_0$) as well as the physical parameters
introduced before: $M$ (total mass), $\eta$ [symmetric mass-ratio
(\ref{eq:symm_massratio})] and $\chi$ [spin combination
(\ref{eq:chi_definition})]. The efficiency of detecting a signal
defined by $t_0, \phi_0, M,  \eta$ and $\chi$
is properly quantified through the fitting factor
\begin{equation}
 \textrm{FF} = \max_{M', \eta' , \chi'}  \mathcal O \Big[
h_1(M',\eta',\chi'), h_2
(M,\eta,\chi) \Big]. \label{eq:FF}
\end{equation}
Note that the maximization with respect to $t_0$ and $\phi_0$
is already included in the definition of the overlap $\mathcal O$,
see (\ref{eq:Ov}).

The accuracy threshold for detection we quoted before is indeed
defined including this additional maximization, i.e., in terms of
\begin{equation}
 \mathcal M_{\rm FF} = 1 - \textrm{FF}~. \label{eq:MMFF}
\end{equation}
If a waveform family $\{ h_1 \}$ satisfies $\mathcal M_{\rm FF} (h_1,
h_2) < \mathcal M_{\rm max}$ (with sufficiently small $\mathcal M_{\rm
max}$) then it is said to be \emph{effectual}
in the detection of the target signal $h_2$ \cite{Damour:1997ub}. The
results in
Sec.~\ref{sec:hybMM} are only a lower bound on this effectualness.

The accuracy requirements for parameter estimation are naturally more
demanding than those for detection. In the recent literature
\cite{Lindblom:2008cm,Santamaria:2010yb,Damour:2010zb,
MacDonald:2011ne} the \emph{faithfulness} of waveforms was usually
defined by
the criterion~(\ref{eq:indist_h}) (without optimization with respect
to
physical parameters), thereby demanding that the maximal information
can be extracted from the data without being restricted by the model
itself. Here, however, we want to understand faithfulness in the
original sense introduced in \cite{Damour:1997ub} that is based on
the difference of the target waveform parameter $\lambda$ with
the recovered model parameter $\bar \lambda$ for which (\ref{eq:FF}) is
maximal. If this bias $\Delta \lambda = \bar \lambda - \lambda$ is small enough, we
can still accept the waveform model family as sufficiently accurate,
even
for parameter estimation. Therefore, by analyzing $\mathcal M_{\rm
FF}$ and the corresponding parameters we can sensibly make
analogous conclusions as before, but based on the actual optimization
strategy that is employed in current template-based GW searches.

Because of the additional freedom of varying physical parameters
we now have to calculate the \emph{ambiguity function}
\begin{equation}
 \mathcal A (\bm {\lambda}', \bm{ \lambda})
 = \mathcal O \left[ h_1 (\bm {\lambda}'), h_2(\bm{\lambda})
\right] \label{eq:ambiguity}
\end{equation}
between hybrids constructed from the same set of NR waveforms but
members of different PN approximants. It depends on the parameters
of the waveforms, $\bm \lambda'$ and $\bm{\lambda}$, as well as
the waveform models themselves.

Since the phase difference above $M\omega_m$ in the overlap integral
(\ref{eq:Ov}) does not vanish generally for $\bm \lambda' \neq \bm{
\lambda}$, we have to slightly modify the algorithm presented in
Sec.~\ref{sec:hybMM_general}. In particular, we now need an estimate
of how small changes in physical parameters affect the phase
difference in the assumed NR regime. (The PN regime is affected as
well, but there is no qualitative difference to the PN comparison
incorporated before.) One possible strategy to quantify phase changes
along variable physical parameters is to perform a number of numerical
simulations and interpolate between the data obtained. Depending on
the density of samples in the $\eta$ and $\chi$ directions
(the scaling with $M$ is given trivially by a single simulation), such
a procedure can be very time- and resource-consuming. However, the
phenomenological fittings performed in
\cite{Ajith:2007qp,Ajith:2007kx,Ajith:2009bn,Santamaria:2010yb} have
utilized exactly this type of interpolation, and we conveniently use
the result of \cite{Santamaria:2010yb} here because the fitting there
is localized to frequencies close to and in the NR regime.

Finally, to ensure the proper relative alignment, our algorithm to
calculate $\mathcal A$ for arbitrary (in practice small) variations
in all parameters is to match different PN approximants to a
phenomenological waveform (phase and amplitude) that is used above $M
\omega_m$ resulting in a hybrid $\tilde h
(f;M,\eta,\chi,t_0,\phi_0)$. 

\begin{figure}
 \centering
\includegraphics[width=0.45\columnwidth]{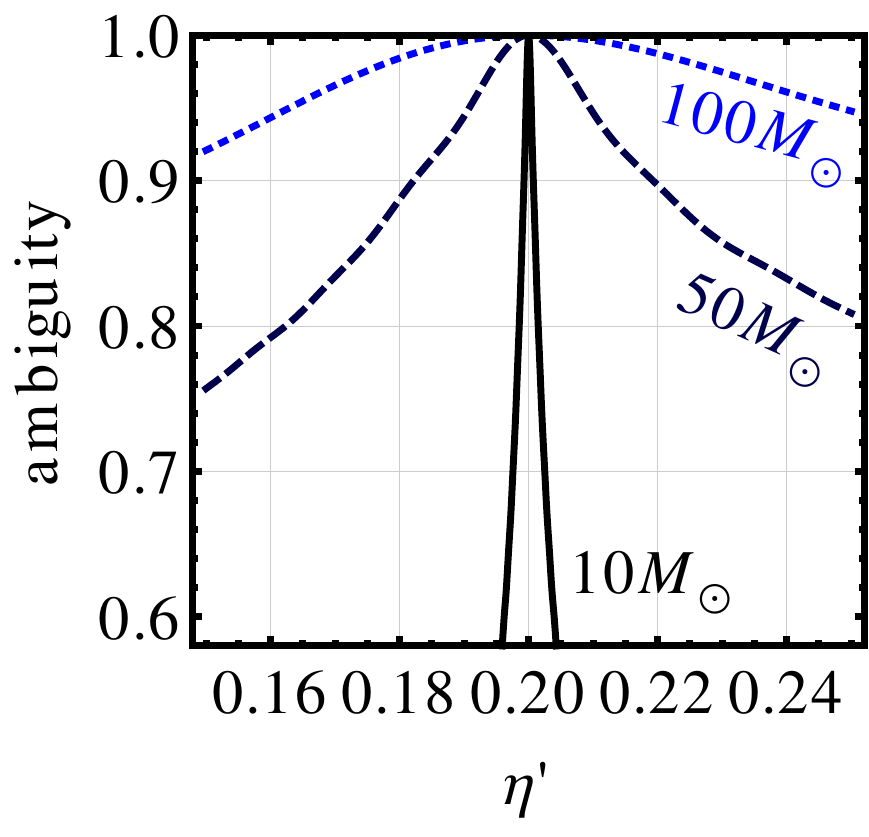} ~~
\includegraphics[width=0.45\columnwidth]{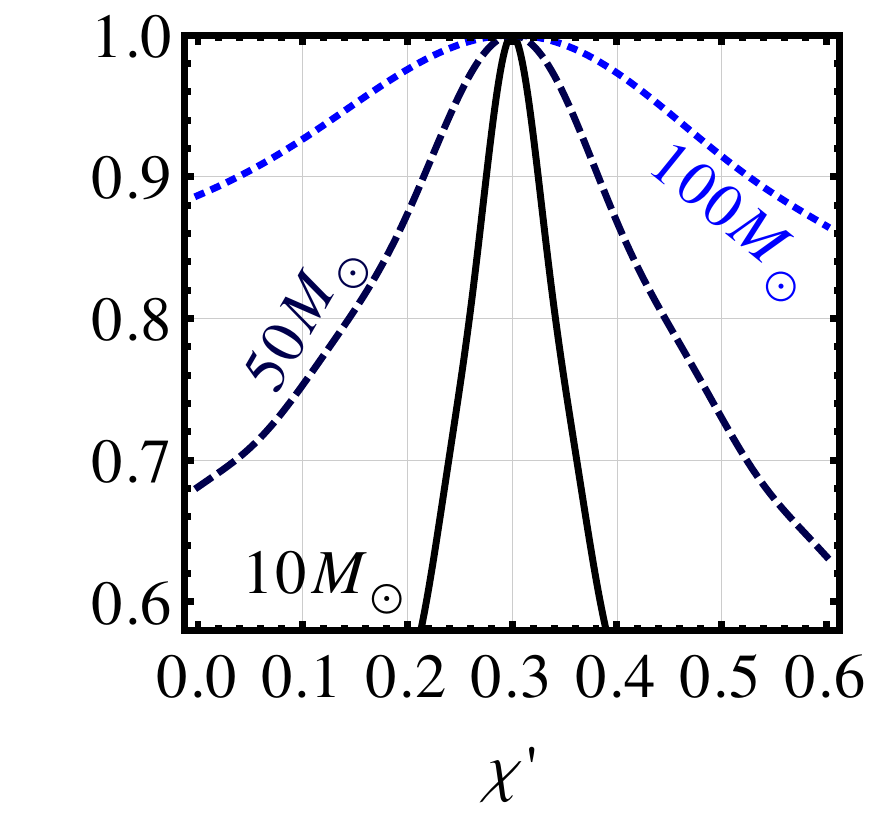} \\[6pt]
\includegraphics[width=0.72\columnwidth]{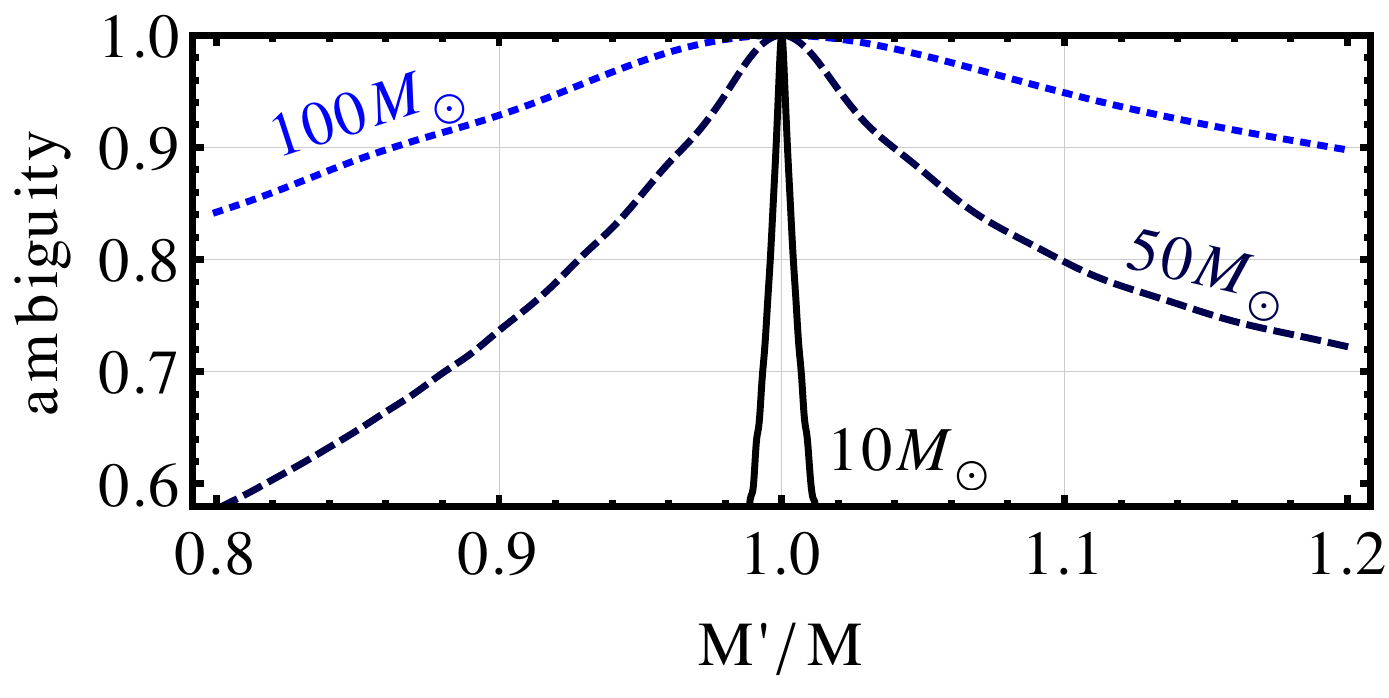} 
\caption{The ambiguity function (\ref{eq:ambiguity}) between two
phenomenological waveforms \cite{Santamaria:2010yb}, where $h_2$
is fixed with $\eta = 0.2$, $\chi = 0.3$ and the total
mass as indicated in the plots. The parameters of $h_1$ are varied
individually while the others are kept constant at the values of $h_2$, respectively.}
\label{fig:ambiguity}
\end{figure}

Let us highlight that although we are now building PN+phenomenological
hybrids
our analysis is not assessing how accurate individual waveforms
describe the entire coalescence process. Note for instance that we
could have introduced this hybridization concept already in the
previous section, but, as we have shown, the phase above the matching
frequency did not enter the overlap calculation. Similarly now, we
use the phenomenological phase description merely to model the $M$-,
$\eta$- and $\chi$-dependence at higher frequencies.
Figure~\ref{fig:ambiguity}
illustrates what kind of information we are using by plotting
slices of the
ambiguity function of the phenomenological model with itself for the
case $\eta = 0.2$ (mass-ratio $\approx 2.6$), $\chi = 0.3$ and
$M/M_\odot \in
\{10,50,100\}$. In Sec.~\ref{sec:hybMM} we only exploited $\mathcal A
= 1$ for $\bm \lambda' = \bm{\lambda}$ whereas now we need an
estimate of the shape of $\mathcal A$ also for $\bm \lambda' \neq
\bm{\lambda}$ (although for small $\vert \bm \lambda' - \bm{
\lambda} \vert$).

We can make two immediate observations from
Fig.~\ref{fig:ambiguity}. Especially for small masses we see that
relatively small changes in, for instance, symmetric mass-ratio or
total mass (the other parameters are kept constant, respectively)
modify the waveform considerably, so that the high mismatches for
equal parameters (reported, e.g., in Figs.~\ref{fig:hybEst} and
\ref{fig:MMContours}) could
potentially be reduced drastically by only small variations in the
physical parameters of one model waveform. Although the formal
criterion (\ref{eq:indist_h}) for faithfulness (or better
indistinguishability) failed, the fitting factor could still be
extremely close to unity with a minimal bias in the parameters. The
second
interesting observation from Fig.~\ref{fig:ambiguity} is that the
width around the maximum of the ambiguity function increases towards
higher masses so that a comparison of two waveforms is increasingly
insensitive to parameter changes at higher frequencies. This in
turn endorses our assumption that the fitting factors and biases we
shall calculate are dominated by PN effects (and not the choice of 
data above $M \omega_m$) for
small masses, where
the accuracy requirements turned out to be hardest to satisfy.

\subsection{Comparison with previous results}

\begin{figure}
 \centering
\includegraphics[width=0.87\columnwidth]{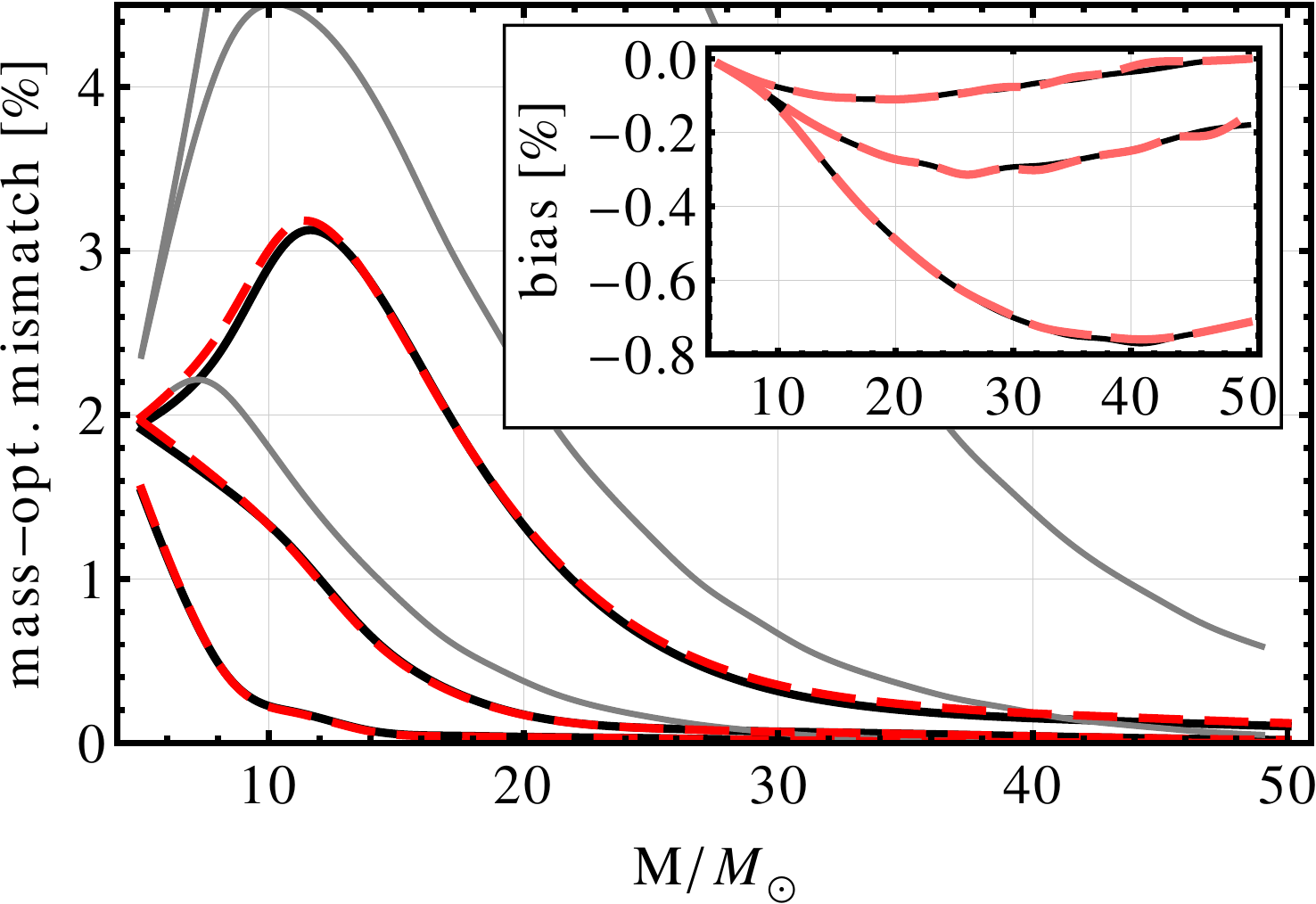}
\caption{The mass-optimized mismatch between
equal-mass, nonspinning TaylorT1/TaylorT4+NR hybrids (black solid
lines) compared to our NR-free estimate (dashed lines). The matching
frequencies are $M \omega_m \in \{ 0.04,0.06,0.08
\}$ from bottom to top. The gray lines show the results of
nonoptimized mismatches for comparison, see also
Fig.~\ref{fig:hybEst}. The inset illustrates the relative bias in the
total mass (matching frequencies in reverse order).}
\label{fig:mass-opt} 
\end{figure}

Before exploring fitting factors across the parameter
space, let us present two examples that illustrate the
general conclusions we shall draw in this paper. We first come back
to the canonical equal-mass, nonspinning case and the
TaylorT1/TaylorT4 comparison that was employed before (see
Fig.~\ref{fig:hybEst} and \cite{Hannam:2010ky}). To test the validity
of our approach we again compare our estimate to hybrids constructed
with actual NR data (matched at $M \omega_m \in \{ 0.04,0.06,0.08
\}$, respectively). Because of the unavailability of NR data with
arbitrary $\eta$ and $\chi$, we
for now only maximize with respect to the total mass $M$. 
Note that the results shown in 
Fig.~\ref{fig:mass-opt} fully agree with the analysis of
Hannam \emph{et al.}~\cite{Hannam:2010ky} (see Fig.~6 therein). They
not only confirm that
our combination of PN
and phenomenological data accurately predicts the disagreement of
the ``true'' PN+NR hybrids, one can also observe the striking
improvement when the additional maximization with respect to $M$ is
taken into account. The peak mismatch without optimization was
approximately 8.8\%, 4.5\% or 2.2\%, depending on $M \omega_m$. With
mass optimization we instead find $\mathcal M_{\rm FF} < 3.2\%$,
2.0\% and 1.5\%, respectively. The relative bias in the total mass,
$(\bar M - M)/M$, is always less than 0.8\% and the earlier the
matching is performed the smaller the bias becomes. 

\begin{figure*}
\includegraphics[width=0.42\textwidth]{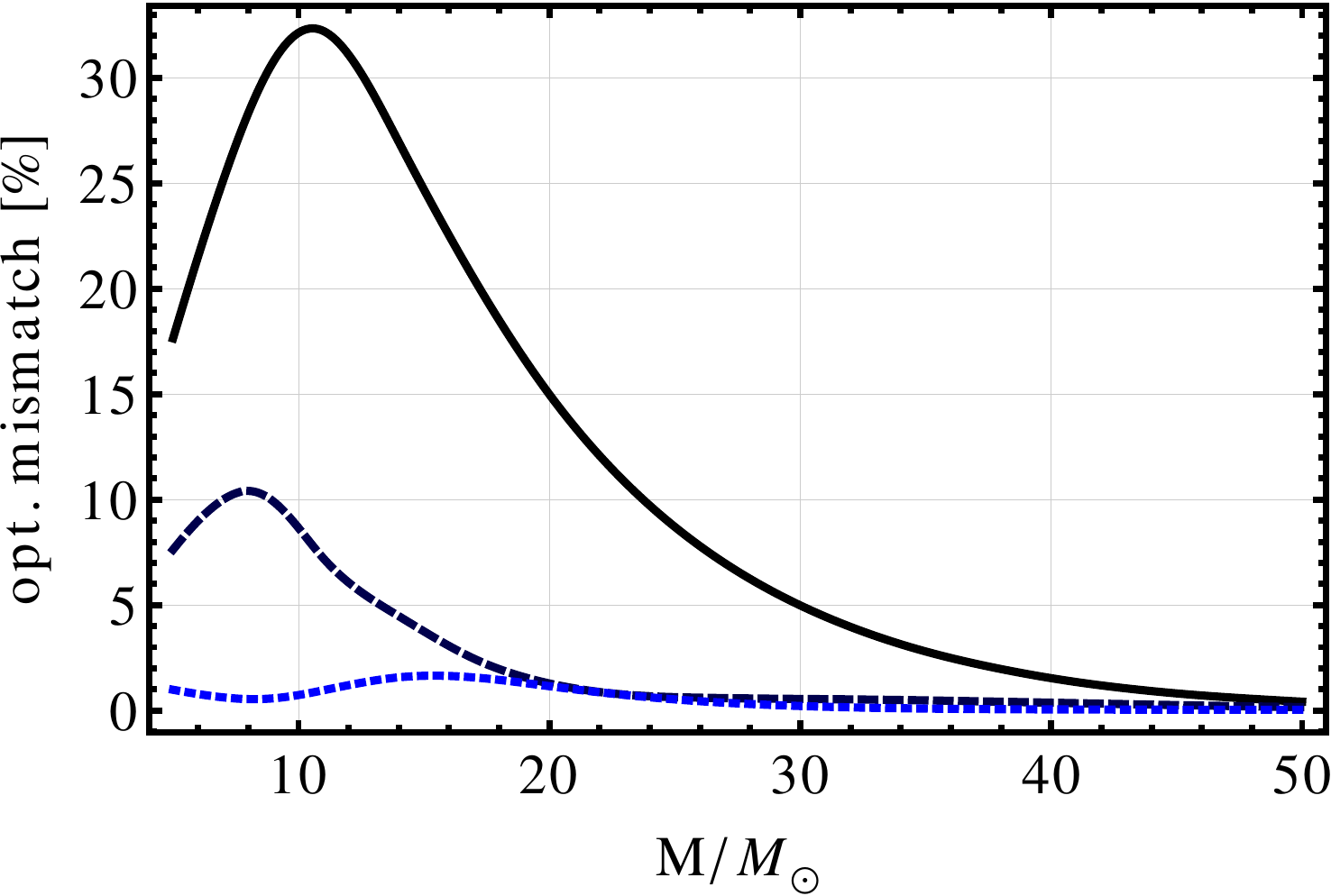}\hfil
\includegraphics[width=0.42\textwidth]{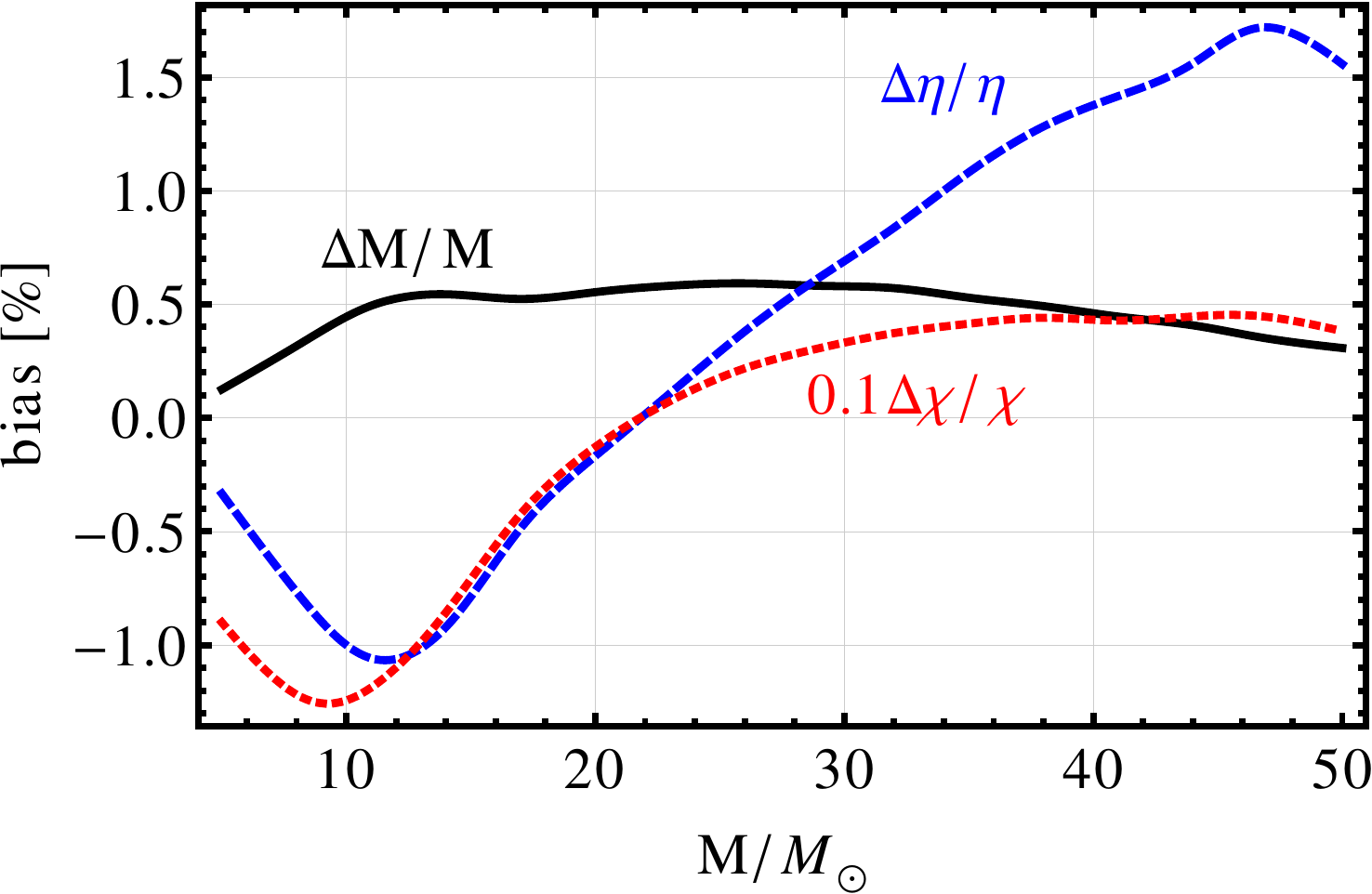} 
\caption{Mismatch between TaylorT1- and TaylorF2-based waveforms
for mass ratio 4:1, $\chi = 0.5$ and matching frequency $M\omega_m =
0.06$ (left panel). The mismatch is not optimized (top solid line), mass-optimized
(dashed line) or optimized with respect to all physical parameters of
the TaylorF2-based model (lowest dotted line). The bias in the
parameters are provided on the right panel.}
\label{fig:q4s5full-opt}
\end{figure*}

A subsequent question that has not been answered so far is to what
extent further optimizations, say along the symmetric mass-ratio
and the spin(s) of the model system, improve the agreement between the
waveform families even more. Full fitting factor calculations are
commonly used to compare waveform models (see, e.g.,
\cite{Buonanno:2009zt,Damour:2010zb}), but they have not been employed
in the context of hybrid waveforms and studies of the required length
of numerical waveforms. 
Reference \cite{Hannam:2010ky} only applied a crude estimation of the
effect an additional mass-ratio optimization has, and 
 concluded that a (total) mass-optimization alone serves as
a sufficient assessment of the full fitting factor. We now find that this 
conclusion was incorrect. 
We illustrate the effect of further optimizations through the
comparison of TaylorT1- and TaylorF2-based waveforms (matched at
$M\omega_m = 0.06$) in Fig.~\ref{fig:q4s5full-opt}.
The TaylorT1 target signal is fixed as a system with mass-ratio 4:1
and spin $\chi = 0.5$, a point in parameter space that clearly fails
all accuracy requirements when looking at Fig.~\ref{fig:MMContours}.
By maximizing with respect to $M$, however, the maximal mismatch
drops from 32.2\% to 10.4\%. Varying all three considered physical
parameters finally yields a curve with $\mathcal M_{\rm FF} \approx
1.6\%$ at maximum, making the TaylorF2-based family 
accurate enough for detection. The relative bias in the parameters are
less than 1\% for $M$, of the order of 1\% for $\eta$ and $\lesssim
10\%$
for $\chi$. 

Note that a faithfulness analysis, as in Sec.~\ref{sec:hybMM} and 
\cite{MacDonald:2011ne,Boyle:2011dy}, would conclude that NR waveforms with many hundreds of cycles
are necessary to produce hybrids (and consequently waveform models) that
are sufficient for parameter estimation purposes. Here we see that waveforms
that we might at first sight regard as far too inaccurate, in fact may
yield 
relatively small parameter biases when embedded in a waveform family.

The optimization algorithm with respect to physical parameters is
computationally more challenging than maximizing the inner product
with respect to $t_0$ and $\phi_0$ only. For each set of test
parameters $(\eta, \chi)$ we have to construct a new waveform. Since
TaylorF2 is an analytical closed-form PN description that is fast to
evaluate
and our matching to the phenomenological model is performed directly
in Fourier space \cite{Santamaria:2010yb} we only consider
TaylorF2-hybrids as test waveforms $h_1$. For the fixed target
waveforms $h_2$ we chose to employ the TaylorT1
approximant, because it was shown in \cite{Hannam:2010ec} that its
(dis)agreement to premerger NR data is most robust over the
considered parameter space and \cite{Boyle:2011dy} noted that a
maximal uncertainty estimate involves comparing to TaylorT1-inspirals.

Starting with equal parameters $\bm{\lambda}' = \bm{
\lambda}$,
we search for the nearest local maximum of the overlap $\mathcal O
(h_1, h_2)$ by varying $\bm {\lambda}'$ along the gradient of the
overlap. Thus, we ensure a quickly converging improvement after a
relatively small number of iterations. The results we present,
however, do not take into account the entire distribution of the
ambiguity function and are still only a lower bound on the fitting
factor. Given the tremendous decrease in mismatch for relatively small
changes in physical parameters we argue nevertheless that this local
extremum should serve as a reasonable estimate of the error one has
to assume in terms of the fitting factor. 

\begin{figure}
 \centering
\includegraphics[width=0.9\columnwidth]{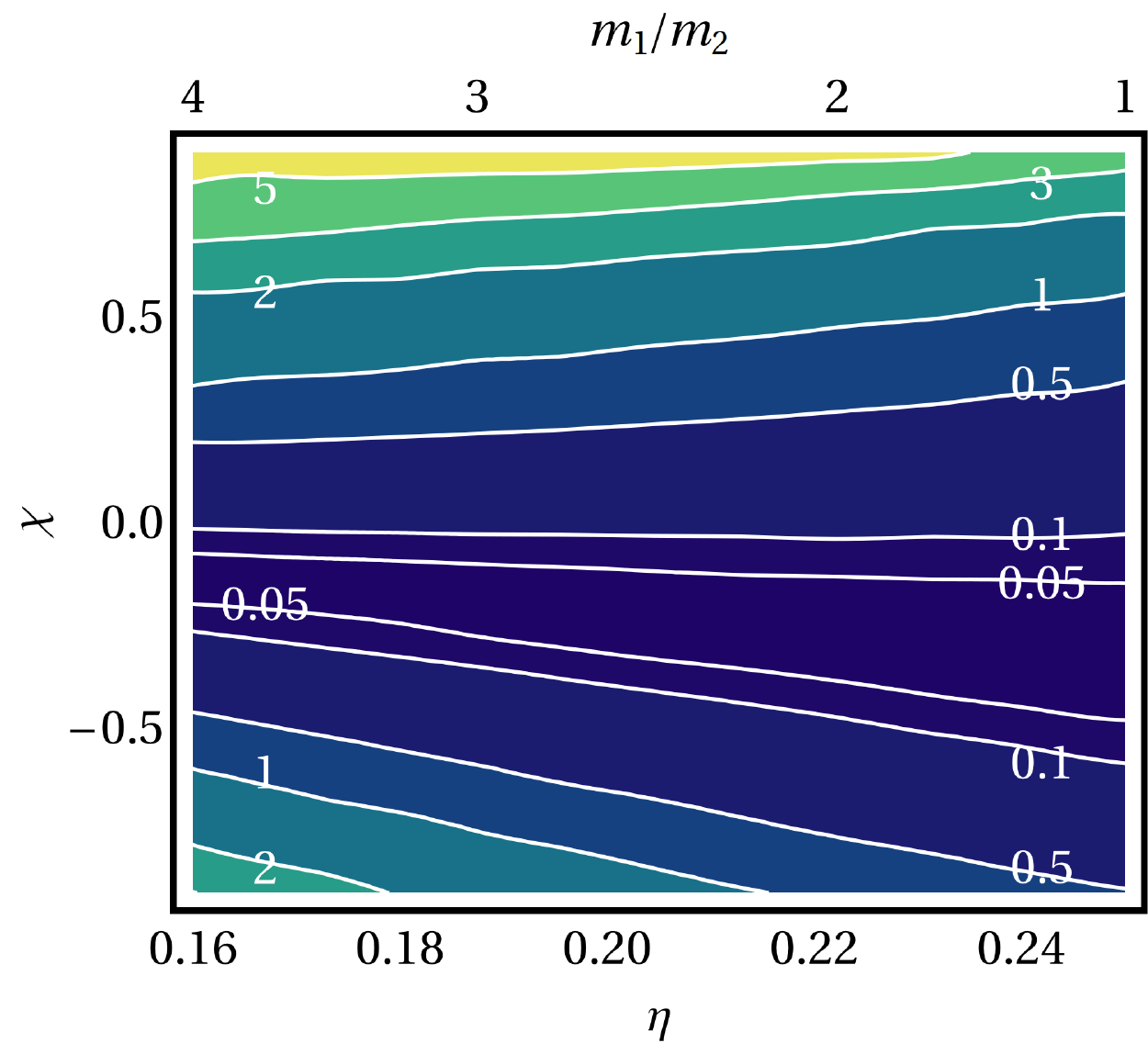}
\caption{The maximum of the optimized mismatch (in \%) for
hybrids constructed either with TaylorT1 (target signal) or TaylorF2
(template signal) and a matching frequency of $M\omega_m = 0.06$.}
\label{fig:FFContour}
\end{figure}

We repeated the exploration of the parameter space with a study
similar to the one presented in Fig.~\ref{fig:MMContours}. The
matching frequency is again fixed at $M \omega_m = 0.06$ and we
calculate $\mathcal M_{\rm FF}$, Eq.~(\ref{eq:MMFF}), for masses $5
M_\odot \leq M \leq 20 M_\odot$. We checked that the mismatch decreases towards the boundaries of this interval, so that the enclosed maximum can indeed be regarded as the global extremum.
After performing this maximization of the mismatch with respect to $M$ for
fixed $(\eta, \chi)$, we present our results as a
contour plot in Fig.~\ref{fig:FFContour}. The structure is very
similar to the pattern of the nonoptimized mismatch, cf. the right
panel of Fig.~\ref{fig:MMContours}. The obvious difference is,
however, that calculating the detection-relevant
quantity $\mathcal M_{\rm FF}$ instead of the diagonal mismatch $1 -
\mathcal A(\bm{\lambda},  \bm{\lambda})$ results in numbers that are
$\sim 10$ times less than what was considered before as error
estimates.

This allows for very different conclusions: Even a moderate matching
frequency like the one considered here leads to hybrids that are
accurate enough for detection in a large portion of the parameter
space. Simulating NR waveforms with few ($< 10$) orbits should hence 
be good enough for many applications considering systems with moderate
spins and mass-ratios. Although this is a very broad statement, it is
clearly distinct from previous analyses 
\cite{Damour:2010zb,MacDonald:2011ne,Boyle:2011dy} that concluded
\emph{much longer} NR waveforms are needed to sensibly connect them
to standard PN approximants. 

Of course, Fig.~\ref{fig:FFContour} only shows the optimal agreement
between the two considered waveform families and one might fear that
the difference between simulated and recovered parameters is large in
some parts of the parameter space. However, as anticipated by
Fig.~\ref{fig:q4s5full-opt}, the bias in total mass and symmetric
mass-ratio are small, approximately  $\pm 1\%$ and $\pm 1.5\%$
at most, respectively. The spin parameter $\chi$ is uncertain by
$-0.15 \leq \Delta \chi \leq 0.05$. A deeper analysis of these
biases is beyond the scope of this paper and results are likely
more model-dependent than the general conclusions we present here. 

For
completeness, we note that for increasing values of the simulated
spin, 
$\Delta \eta$ and $\Delta \chi$ generally
decrease from positive to negative values, $\Delta M$ increases at
the same time. This correlation is expected from the form of the PN
expansion, where modifications of $M$ can be compensated at lowest
order by changing $\eta$ inversely. Studies of PN approximants in
\cite{Buonanno:2009zt} show similar tendencies, although the biases
reported there are considerably higher due to the absence of a common
NR part at high frequencies. 
The same holds for the comparison of complete models (including
uncertainties in the NR regime)~\cite{Damour:2010zb}.
The modeling biases we find should be compared to statistical errors
of full waveform families. In the case of the nonspinning
phenomenological model \cite{Ajith:2007kx} a Fisher matrix 
study as well as Monte-Carlo simulations were presented in
\cite{Ajith:2009fz}, and the uncertainties found for Advanced LIGO and
signals of SNR 10 are $\Delta M/M \lesssim 3\%$ and $\Delta \eta/\eta
\lesssim 8\%$ ($M < 100M_\odot$). These values are of the same order
of magnitude as our results, and we take this as an indication that
modeling errors do not vastly dominate the parameter estimation
uncertainty. However, further studies are underway \cite{Robinson} to
determine statistical errors for spinning waveform models.

\subsection{Model accuracy for spinning systems}

These new results constitute much brighter prospects for currently
feasible NR simulations than the conclusions drawn in
Sec.~\ref{sec:hybMM} and
\cite{Hannam:2010ky,Damour:2010zb,MacDonald:2011ne,Boyle:2011dy}. In
certain parts of the parameter space,
however, the mismatch error presented in Fig.~\ref{fig:FFContour} is
still too high, particularly if one keeps in mind that gaining
sensitivity of GW detectors is extremely difficult on the hardware
side and theoretical considerations should reduce this sensitivity as
little as possible \cite{Damour:2010zb}. Therefore,
$\mathcal M_{\rm FF} > 3\%$ for highly spinning systems should be
improved by considering lower matching frequencies. Equally important
is the question of whether numerical simulations for systems with
moderate spins and mass ratios can be considerably shorter than $M
\omega_m = 0.06$ which we assumed so far.

\begin{figure}
 \centering
\includegraphics[width=0.9\columnwidth]{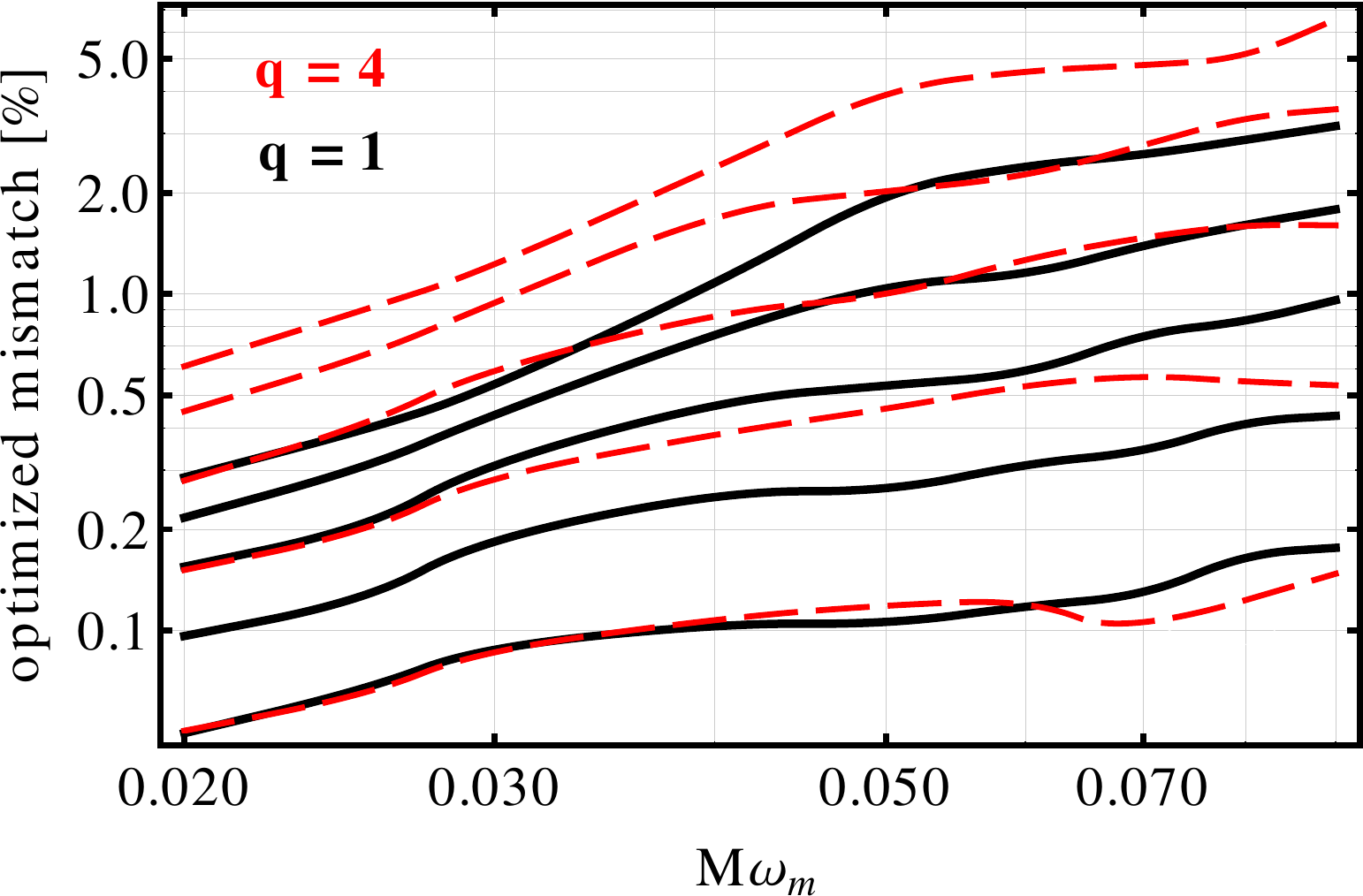}
\caption{The fully optimized mismatch $\mathcal M_{\rm FF}$ as a
function of the matching frequency $M \omega_m$ for equal-mass
systems (solid lines) and mass-ratio 4:1 (dashed lines). The
considered spins in each case are $\chi \in \{ 0, 0.2, 0.4, 0.6, 0.8
\}$ from bottom to top.}
\label{fig:FFomega}
\end{figure}

In Fig.~\ref{fig:FFomega} we analyze the dependence of the mismatch
error by showing the maximum of $\mathcal M_{\rm FF}$ as a function of
$M \omega_m$. We consider equal
masses and mass-ratio 4:1 with spins $\chi \in \{
0,0.2,0.4,0.6,0.8\}$ in each case. Note that we do not include negative values of $\chi$ here, because the fact that the mismatch
error for $\chi <0 $ is smaller and not monotonic in $\chi$ (see Fig.~\ref{fig:FFContour}) is likely an artifact of our choice of
PN approximants (recall the obvious differences in
Fig.~\ref{fig:MMContours}). 
 As expected, Fig.~\ref{fig:FFomega} illustrates that reducing the matching
frequency, e.g., from $M \omega_m = 0.08$ to $M \omega_m = 0.02$, leads
to an improvement in mismatch by a factor of 2 to 10, depending on the
spin. 

Larger values of the spin generally yield larger
mismatches which in turn leads to stronger requirements for $M
\omega_m$, assuming a given accuracy goal. This is unfortunate because
the orbital hangup configuration of positive aligned spins
decelerates the frequency evolution in the inspiral of the binary,
demanding even longer simulations for a given frequency range.

\begin{table}
\begin{ruledtabular}
 \begin{tabular}{|c|c|c|}
orbits & equal-mass & mass-ratio 4:1 \\ \hline
\multirow{2}{*}{5} & \emph{1.5\%:} $-0.76 < \chi
< 0.60 $ & \emph{3.0\%:} $-0.95 < \chi
< 0.55$\\
& \emph{0.5\%:} $-0.37 < \chi
< 0.31$ &  \emph{1.5\%:} $-0.52 < \chi
< 0.39$  \\ \hline
\multirow{2}{*}{10} & \emph{1.5\%:} $-1.00 \leq
\chi
< 0.70$ &  \emph{3.0\%:} $-1.00 \leq \chi
< 0.68$ \\
& \emph{0.5\%:}  $-0.45 < \chi
< 0.39$ &  \emph{1.5\%:} $-0.56 < \chi
< 0.48$ \\ \hline
\multirow{2}{*}{20} & \emph{0.5\%:} $-0.97 < \chi
< 0.57$ & \emph{3.0\%:} $-1.00 \leq \chi
< 0.79 $ \\
& \emph{0.2\%:} $-0.28 < \chi
< 0.22$  & \emph{1.5\%:} $-0.92 < \chi
< 0.55$
 \end{tabular}
\end{ruledtabular}
\caption{Range in spin parameter $\chi$ where a given accuracy
requirement ($\mathcal M_{\rm FF} < 3\%, 1.5\%, 0.5\%$ or $0.2\%$) is
fulfilled. Each row specifies the assumed number of orbits before
merger for the NR waveform ($=$ number of GW cycles divided by 2).}
\label{tab:NRlength}
\end{table}

As such extremely long NR waveforms may not be available in the near
future (including the Advanced LIGO era), we continue with a slightly
different application of our results:
How reliable is a set of complete waveforms constructed with standard
PN approximants and NR simulations covering 5 (10, 20) orbits before
merger (i.e., 10, 20 or 40 GW cycles prior to the maximum of $\vert h
(t)\vert$)? To quantify these uncertainties we have to combine an
estimate of the minimal matching frequency allowed by such
NR waveforms with the resulting mismatch error from
Fig.~\ref{fig:FFomega}. We calculate the first from the inverse
Fourier transform of the phenomenological model
\cite{Santamaria:2010yb} and the time derivative of the 
phase, $M \omega_m \approx d \arg h(t_n) /dt$, where $\arg (t_n) =
\arg h(t_{\rm max}) - n \, 2\pi$ ($n = 10, 20, 40$, respectively), and
$t_{\rm max}$ is the time of the maximum amplitude $\vert h \vert$.
This spin- and $\eta$-dependent value is then taken into the results
presented with Fig.~\ref{fig:FFomega} to estimate $\mathcal M_{\rm
FF}$ for each configuration. Note that we use a more pessimistic error
estimate for antialigned spins ($\chi <0$) by assuming the mismatches
of $\vert \chi \vert$ due to the reasons discussed above.

One kind of possible conclusion one can then draw is summarized in
Table~\ref{tab:NRlength} for equal masses and mass-ratio 4:1.
Given an accuracy goal (which we take as either 3\%, 1.5\%, 0.5\% or
0.2\%) we provide the range of spins in which hybrids with the
specified number of NR orbits fulfill this goal. Note that the
asymmetry in the spin parameter is only caused by the different
matching frequencies waveforms with constant length permit. Again, we
can very clearly see that even relatively short waveform are good
enough for detection. In fact, mismatches of 0.5\% are below the
noise level for SNR 10, and differences of 0.2\% are indistinguishable for
SNR $\lesssim 16$ according to Eq.~(\ref{eq:indist_M}). However, one
can also see from Table~\ref{tab:NRlength} that doubling the number
of orbits does not enlarge the accuracy range dramatically in many
cases, although such simulations would take far more computer power
and time.

\subsection{Nonspinning unequal-mass systems}

So far, we refrained from explicitly calculating mismatches for
mass-ratios $>$ 4:1 here because our underlying phenomenological model
was only calibrated to numerical simulations with mass-ratios $\leq$
4:1. Pushing the model beyond these values would add another
uncertainty in addition to the way we estimate PN errors already, and
more elaborate studies (possible including different models such as
\cite{Ajith:2009bn} and variants of EOBNR) are needed to
reach sound conclusions.

\begin{figure}
 \centering
\includegraphics[width=0.9\columnwidth]{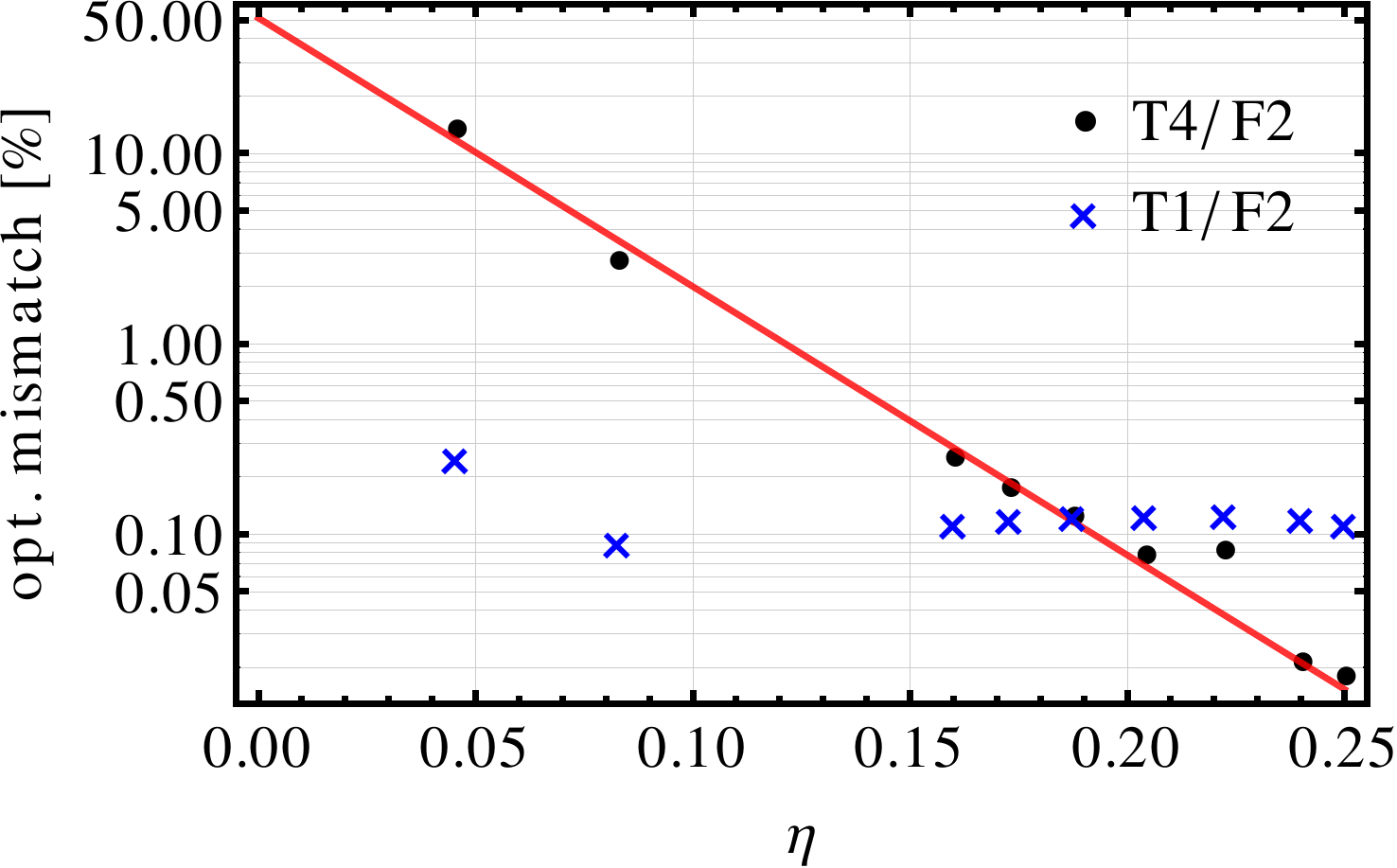}
\caption{The fully optimized mismatch of nonspinning target signals
employing either the TaylorT1 or TaylorT4 approximant with model
waveforms constructed with TaylorF2 inspirals. The assumed matching
frequency is always $M \omega_m = 0.06$. The bias in parameters is
$\vert \Delta M \vert/ M \lesssim 0.6\%$ ($0.16\%$), $\vert \Delta
\eta \vert/ \eta \lesssim 1.0\%$ ($0.3\%$), $\vert \Delta
\chi\vert < 0.04$ ($0.017$), where the values in brackets indicate
the restriction to $q \leq 4$.}
\label{fig:nonsp_maxMM}
\end{figure}

Nevertheless, numerical simulations of higher mass ratios are
potentially interesting, and we shall try to estimate their
reliability on the basis of our (extrapolated) knowledge here. We
restrict this study, however, to nonspinning target signals. These are
the systems where we do not expect the PN errors to drop significantly on
the timescale of Advanced LIGO (in contrast to spinning binaries, where
higher-order PN terms may well be calculated in the next few years). 
We find that the agreement between TaylorT1- and
TaylorF2-based hybrids is exceptionally good along $\chi = 0$ (see
Figs.~\ref{fig:FFContour} and \ref{fig:nonsp_maxMM}). In contrast,
the TaylorT4/TaylorF2 uncertainty increases towards higher
mass-ratios (smaller values of $\eta$) as we would expect from the
form of the PN expansion. Therefore, we shall conservatively base our statements on
comparing TaylorT4 and F2 approximants in this section. 

To illustrate our argument, we plot in Fig.~\ref{fig:nonsp_maxMM} the
maximum of the fully optimized (i.e., with respect to $M$, $\eta$ and
$\chi$) mismatches between TaylorF2 and either TaylorT1 or TaylorT4
hybrids, all matched to fictitious NR data at $M \omega_m = 0.06$. The
fixed target parameters are chosen as $\chi = 0$ with the mass ratio
$q$ varying from 1 to 4 in steps of 0.5 as well as $q = 10$ and $q =
20$. While the comparison with TaylorT1 yields weakly $\eta$-dependent
mismatches below 0.3\%, TaylorT4 target signals exhibit a steeply
increasing divergence from the model signals towards higher
mass ratios. Its approximately exponential behavior is well described
by the following fitting formula
\begin{equation}
 \log_{10} \mathcal M_{\rm FF} \approx -0.29 - 14.1 \eta 
\label{eq:nonsp_MM_fit}
\end{equation}
which is included as a straight line in Fig.~\ref{fig:nonsp_maxMM}. A
conservative estimate of the general model uncertainty would be the
maximum of both data series for each $\eta$, i.e.,
(\ref{eq:nonsp_MM_fit}) for small $\eta$ and roughly constant
$\mathcal M_{\rm FF} \approx 0.12\%$ for $\eta > 0.1866$ ($q < 3$).

Evidently, a
matching frequency of $M \omega_m = 0.06$ is only good enough for
$\eta > 0.081$ ($q < 10.2$) if a mismatch of at most 3\% is
tolerated. Again, reducing the matching frequency helps to increase
the accuracy of the final waveform, and we systematically analyze how
useful numerical simulations of 5, 10 or 20 orbits before merger are
in the
nonspinning unequal-mass regime. For that, we calculate $\max_M
\mathcal M_{\rm FF}$ as a function of the matching frequency and the
symmetric mass ratio, similar to what was done for 
Fig.~\ref{fig:FFomega}. The matching frequency is then converted to
orbits before merger as explained in the previous section. 

\begin{table}
\begin{ruledtabular}
 \begin{tabular}{|c|rl|ll|}
orbits & \multicolumn{2}{c|}{mass-ratio} &\multicolumn{2}{c|}{$ q =
20$} \\ \hline
\multirow{2}{*}{5} &  \emph{3.0\%:} & $q < 8.9$ &
\multicolumn{2}{c|}{$\max_M \mathcal M_{\rm FF} \approx 15\% ~~~ (19
M_\odot)$} \\ 
 & \emph{1.5\%:} & $q < 6.8$ & $21 M_\odot: 12\%,$ & $ 63 M_\odot
: 0.3\%$ \\ \hline
\multirow{2}{*}{10} & \emph{3.0\%:} & $q < 11.4$ & 
\multicolumn{2}{c|}{$\max_M \mathcal M_{\rm FF} \approx 8.2\% ~~~ (13
M_\odot)$} \\
 & \emph{1.5\%:} & $q < 8.6$ & $21 M_\odot: 3.0\%,$ & $ 63 M_\odot
: 1.6 \times 10^{-5}$  \\ \hline
\multirow{2}{*}{20} & \emph{3.0\%:} & $q < 14.8$ & 
\multicolumn{2}{c|}{$\max_M \mathcal M_{\rm FF} \approx 5.7\% ~~~ (11
M_\odot)$} \\
 & \emph{1.5\%:} & $q < 10.7$ & $21 M_\odot: 0.8\%,$ & $ 63 M_\odot
: 6.4 \times 10^{-6}$ \\ 
 \end{tabular}
\end{ruledtabular}
\caption{
Accuracy of nonspinning hybrid waveforms, based on combining
PN TaylorT4 or TaylorF2 data with NR waveforms of specified length
(defined by the number of orbits before merge $=$ number of GW cycles
divided by 2). \emph{Left column:} Range in mass-ratio where a given
accuracy
requirement ($\max_M \mathcal M_{\rm FF} < 3\%$ or  $1.5\%$) is
fulfilled. 
\emph{Right column:} Mismatch error for $q=20$, both at maximum of all
masses (location indicated in parentheses) and at astrophysically
motivated minimal values of the total mass (see text).}
\label{tab:uneqMassOrbits}
\end{table}

In Table~\ref{tab:uneqMassOrbits} we present our results in analogy
to Table~\ref{tab:NRlength}, where we provided the range of the spin
parameter $\chi$ in which the waveform model meets certain accuracy
requirements. Now we complement the picture by restricting ourselves
to the nonspinning case; our error estimates are based on optimized
TaylorT4/TaylorF2 hybrid mismatches, and we present the accuracy range
in terms of the mass ratio. Note that, although only five orbits of
NR data before merger are sufficient for detection for most of
today's standard simulations ($q \lesssim 6$), even the
computationally very challenging goal of 20 orbits before merger is
not enough to reliably model mass ratios as high as 15 or more for
arbitrary total masses of the binary. 

It should be pointed out, however, that we report the worst
disagreement between the considered hybrids in the left column of
Table~\ref{tab:uneqMassOrbits}, i.e., we demand that the assumed
accuracy requirement is satisfied for \emph{all} values of the total
mass. As discussed in \cite{Boyle:2011dy} already, one should rather
understand the mismatch error and the accuracy requirement as
functions of the total mass. After all, binaries with larger total
mass have higher SNR in the detector (for constant distance of the
source). More important for us here is that some of the
considered astrophysical scenarios may not even exist or be extremely
unlikely, and if the modeling error exceeds accuracy thresholds in
these regions, we do not have to bother.

We illustrate this argument with a concrete example: The (fictitious)
waveform of a binary with mass-ratio 20:1 exhibits the largest
uncertainty at total masses less than $20 M_\odot$, depending on the
matching frequency (the values for NR simulations covering 5, 10 or
20 orbits before merger are given in parenthesis in the right column
of Table~\ref{tab:uneqMassOrbits}). If we only consider \emph{black
holes} as objects in the binary and follow observational
\cite{Gelino:2003pr} and theoretical \cite{Belczynski:2001uc} evidence
that their individual masses are $> 3M_\odot$, then the lowest total
mass to consider in our error analysis is instead $63M_\odot$. With
our idealized assumptions, this is a regime where the mismatch drops
monotonically with increasing total mass (due to the dominating amount
of exact high-frequency data), and the maximal uncertainty at
$63M_\odot$ proves to be more than sufficient for detection purposes,
even with only a few NR orbits; see Table~\ref{tab:uneqMassOrbits}. In
this sense, modeling higher mass-ratios is more accurate than
comparable masses, as \cite{Boyle:2011dy} noted already for diagonal
(nonoptimized) mismatches.

One the other hand, one could argue that the smaller object in the
20:1-binary could also be a neutron star. If the companion
is a much heavier black hole, tidal effects are extremely weak
\cite{Pannarale:2011pk} and the plunge is hardly affected from finite
size effects of the neutron star \cite{Shibata:2009cn}. Thus, we may
hope to
accurately capture these systems with a BBH template family as well,
and smaller total masses have to be considered. According to
\cite{Goussard:1997bn},
(proto)neutron stars are expected to have masses $> 1 M_\odot$, which
is in agreement with current observations (see 
\cite{Lattimer:2006xb} for an overview).
Assuming the lower bound of $1 M_\odot$ for the mass of a single
compact object, we consequently have to consider total masses down to
$21M_\odot$ (for $q=20$) which leads to higher modeling uncertainties
in the
waveform. However, as Table~\ref{tab:uneqMassOrbits} shows, 10 NR
orbits before merger would be virtually good enough for detection
purposes, 20 orbits already yield a mismatch of only 0.8\% at
$21M_\odot$. Hence, even the theoretically and numerically difficult
unequal-mass regime may well be modeled with only a few NR orbits,
given the astrophysical expected properties of such systems.

Of course, these astrophysical limitations are highly uncertain, and
the conservative error analyses are the ones presented in
Table~\ref{tab:NRlength} and the left column of
Table~\ref{tab:uneqMassOrbits}. However, given that caveat, we 
conclude that currently feasible numerical simulations are
potentially good enough to model in combination with PN approximants
an important fraction of the parameter space.

\section{Discussion } \label{sec:discussion}

Predicting the GW signature of an inspiraling and merging BBH in
General Relativity is inevitably associated with analytical or
numerical approximations to the full theory, which introduce errors in
the final result $h(t)$ or $\tilde h(f)$. In this paper we estimated
these errors by the distance between two approximate solutions for each
physical configuration. While neglecting uncertainties on the NR side, we
assumed different standard PN approximants in a frequency range up to
the point where the waveform is matched to an NR-based merger and 
ringdown model. 

We quantified the uncertainties by comparing the currently available
3.5PN (spinning contributions up to 2.5PN) versions of TaylorT1,
TaylorT4 and TaylorF2 approximants. Introducing a simple algorithm
that only requires amplitude information beyond the matching
frequency, we first confirmed previous
studies \cite{Hannam:2010ky,MacDonald:2011ne,Boyle:2011dy} that found
 that the mismatch error for fixed physical parameters
greatly exceeds reasonable accuracy requirements, 
assuming  typical NR waveform lengths.  

Instead of demanding extremely long numerical simulations to overcome
this uncertainty in the modeling process, we refined the
understanding of the waveform error by adopting the actual data
analysis strategy of detecting an unknown signal in noise-dominated
interferometer data. In particular, assuming waveform families
instead of individual waveforms naturally redefines
the concept of distance by allowing physical parameters to be varied
in the mismatch calculation.

The results presented in Sec.~\ref{sec:FF} indicate then that the GW
signatures for many astrophysically relevant systems can in fact be
well modeled by
straightforward combinations of standard PN approximants and
currently feasible NR simulations, covering $<10$ orbits before merger. 
The accuracy has not yet
reached a level such that detection and parameter estimation errors 
are limited only by the detector noise for high
SNR events, and the intrinsic uncertainty of BBH models may exceed in
some cases the anticipated deviations caused by non-black holes,
making is impossible to identify them as such.  
Nevertheless, the reported disagreement among different BBH models and
biases in the parameters are certainly tolerable for the first GW
detections that are likely to have low SNRs ($\sim 10$). While this is
true for systems with moderate spins, one has to keep in mind that
even our idealized setting yields mismatch errors for
high values of spins that are of the order of a few percent, which
increase for higher mass ratios. Reducing the matching frequency
poses unrealistic challenges for current NR codes, and either
fundamentally different numerical approaches or advances in PN are
needed to fully control the entire parameter space. 

While the next spin-contributions in PN theory may become available in
the near future to further improve the modeling of spinning systems
(see the recent calculations of higher-order spin-orbit
contributions \cite{Hartung:2011te,Nagar:2011fx,Blanchet:2011zv}),
unequal-mass nonspinning contributions at 4PN order are unlikely to
be calculated with established techniques soon. However, as we
discussed for a binary with mass-ratio 20:1,
astrophysical expectations are that such systems only form
with a high total mass, thereby reducing the impact of PN
uncertainties. Even for 20:1 binaries, our results suggest that NR
simulations of less than 10 orbits are sufficient.

In summary, we found that not single hybrid waveforms, but
rather the embedding in the waveform manifold, results in templates
accurate enough for detection, even with today's limited number of NR
orbits. The uncertainty in physical parameters we had to accept for
this tremendous increase in overlap is rather small, $\sim 1\%$ in
mass and symmetric mass-ratio, and $\sim 0.1$ at most for the spin
parameter $\chi$. For nearly equal-mass systems, the individual masses
of the constituents are then only reliable to 
\begin{equation}
\frac{\Delta m_i  }{m_i} \approx \frac{\Delta M}M +
\sqrt{\frac{\Delta \eta}{\eta}} \sim 10\%~,
\end{equation}
and it has to be decided whether this is good enough for
astrophysical studies. 

Of course, our results rely on a number of assumptions that are 
reasonable in the range where we apply them, but we shall collect
and discuss their generalizations and limitations below. 

First of all, our analyses are meant to provide a general concept of
how
to deal with modeling errors, instead of giving final answers.
Especially, as we emphasized throughout the paper, we do not address
the question of how accurate a particular waveform model is. The
statements formulated here are based on selecting PN
approximants that are compared with each other, and our choices were
made to illustrate the
\emph{order of magnitude} one generally has to assume for our notion
of error. This can be taken as a conservative estimate 
for all currently exsisting combinations of analytical and numerical
relativity, because even a remarkable agreement in the overlapping
region of both approaches does not necessarily diminish the
uncertainty of many ambiguous choices that enter the modeling of
(up to thousands of) GW cycles in the inspiral waveform.
Nevertheless, one should keep in mind that a particular
PN (or EOB)+NR combination can be much closer to the real waveform
than estimated here, as well as the possibility that the PN
ambiguity at consistent 3.5PN order generally underestimates the true
error in the signal description. 

Two further essential assumptions should be noted: We neglect both the
error of the
hybridization procedure and any uncertainties beyond the matching
frequency. Both assumptions are well motivated by previous studies
\cite{Hannam:2009hh,Santamaria:2010yb,Hannam:2010ky,MacDonald:2011ne},
but care has to be taken when generalizing their validity. For
instance, from Fig.~\ref{fig:FFContour} or Table~\ref{tab:NRlength}
one might be tempted to conclude that actually very short NR waveforms
are enough for modeling equal-mass, hardly spinning systems. This is
certainly true from our results if the matching to PN can be done
unambiguously. However, if there are too few cycles to align PN and
NR signals properly, different matching procedures may lead to very
different results. This aspect was not treated here as it can be
checked separately, and it should only affect the resulting waveform
for very short ($<5$ orbits) NR simulations.

The other key assumption, the presence of exact high-frequency data,
implies another important aspect to our results. Not
only do we say that the error of the NR part of the wave is
negligible (an assumption that could easily be dropped if the NR
mismatch becomes significant) we also use waveform families that
directly resemble PN/NR hybrids. In other words, the additional error
that is introduced in the phenomenological fitting and interpolation
process is not taken into account here. Again, this is an error that
can be quantified separately, but it has to be taken into account
when interpreting the comparison of different complete waveform
models, as was done in \cite{Damour:2010zb}. We merely state the fact
here that \emph{in principle} PN+NR combinations constitute
sufficiently accurate target waveforms for the construction of
template families.

This work can be complemented in many different ways. 
One obvious, yet involved extension is the completion of the
parameter space by allowing arbitrary spin orientations that cause
additional precession dynamics. Some steps towards building such
hybrids have taken place already
\cite{Campanelli:2008nk,Schmidt:2010it}, but a deeper understanding of
the waveform structure has to be gained before an extensive error
analysis like the present one can be performed. Similarly, this study
was restricted to the dominant spherical harmonic mode as it is
crucial to understand and quantify the errors here first.
Nevertheless, a final waveform model would have to include higher
modes as well, and the algorithm we presented should be easily adaptable
to these cases. 

Implementing more PN
approximants and repeating our analysis with pairwise comparisons of
various flavors of PN and EOB will help to fully
understand the spread of equivalent descriptions of the inspiral
process. When more contributions to PN expansions become available the
present analysis has to be repeated, hopefully reflecting
the enhanced knowledge of the analytical approximation. This is especially
true for spinning binaries, where calculations of higher-order PN contributions
are expected in the next few years.

Also, work is already underway \cite{Robinson} to extend previous work \cite{Ajith:2009fz} and relate the parameter
uncertainties found in this study to statistical errors that are
inevitably present for signals with a given SNR in the detector. Only
these results will allow for statements about how useful current
waveform constructions are for parameter estimation and if the
uncertainty in recovered parameters is dominated by the detector
noise or the waveform model itself.

Finally and most interestingly, one should address the question of
what kind of physics can be achieved given a certain performance of
complete waveform models and, of course, given real GW detections
with the upcoming generation of interferometers. It will be
particularly important to analyze whether a certain disagreement
between signal and model can be entirely explained by model
uncertainties or if possibly unknown physical effects are the cause.
This study serves as a first step to prepare for those kinds of
questions.

\section*{Acknowledgments}

It is a pleasure to thank P. Ajith, S. Babak, B. Krishnan, F.
Pannarale and E. Robinson for useful discussions. 
F. Ohme thanks Cardiff University for hospitality while some of
this work was carried out. This work was supported in part by
the DLR (Deutsches Zentrum f\"ur Luft-und Raumfahrt) and the IMPRS for
Gravitational Wave Astronomy.
M. Hannam was supported by FWF grant M1178
and Science and Technology Facilities Council grants ST/H008438/1
and ST/I001085/1.
S. Husa was supported by DAAD grant D/07/13385 and
grants FPA-2007-60220 and FPA-2010-16495 from the Spanish Ministry of Science,
and
the Spanish MICINN’s Consolider-Ingenio 2010 Programme under grant 
MultiDark CSD2009-00064.

\bibliography{HybridMismatches}

\end{document}